\documentclass[preprint,12pt]{elsarticle}

\usepackage{graphicx}
\usepackage{amssymb}
\usepackage{amsbsy}
\usepackage{amsmath}
\usepackage{xcolor}
\usepackage{ulem}
\usepackage{soul}
\usepackage{multirow}

\usepackage{lineno}

\makeatletter
\def\ps@pprintTitle{%
 \let\@oddhead\@empty
 \let\@evenhead\@empty
 \def\@oddfoot{}%
 \let\@evenfoot\@oddfoot}
\makeatother

\begin{document}

\begin{frontmatter}


\title{6G Networks: Beyond Shannon Towards Semantic and Goal-Oriented Communications}






\author[label1]{Emilio Calvanese Strinati\corref{cor1}\fnref{label3}}
\address[label1]{CEA Leti, 17 rue des Martyres, 38000 Grenoble, France}

\cortext[cor1]{I am corresponding author}

\ead{emilio.calvanese-strinati@cea.fr}

\author[label5]{Sergio Barbarossa}
\address[label5]{Dept. of Information Eng., Electronics, and Telecomm., Sapienza Univ. of Rome\\
Via Eudossiana 18, 00184, Rome, Italy}
\ead{sergio.barbarossa@uniroma1.it}
\begin{abstract}
The goal of this paper is to promote the idea that including semantic and goal-oriented aspects in future 6G networks can produce a significant leap forward in terms of system effectiveness and sustainability. Semantic communication goes beyond the common Shannon paradigm of guaranteeing the correct reception of each single transmitted bit, irrespective of the meaning conveyed by the transmitted bits. The idea is that, whenever communication occurs to convey meaning or to accomplish a goal, what really matters is the impact that the received bits have on the interpretation of the meaning intended by the transmitter or on the accomplishment of a common goal. Focusing on semantic and goal-oriented aspects, and possibly combining them,  helps to identify the {\it relevant information}, i.e. the information strictly necessary to recover the meaning intended by the transmitter or to accomplish a goal. Combining knowledge representation and reasoning tools with machine learning algorithms paves the way to build semantic learning strategies enabling current machine learning algorithms to achieve better interpretation capabilities and contrast adversarial attacks. 6G semantic networks can bring semantic learning mechanisms at the edge of the network and, at the same time, semantic learning can help 6G networks to improve their efficiency and sustainability.
\end{abstract}


\end{frontmatter}





\section{Introduction}
\label{Sec:Intro}
Even though 5G networks are still at an early deployment stage, they already represent a breakthrough in the design of communication networks, shaped around their ability to provide a single platform enabling a variety of different services, ranging from enhanced Mobile BroadBand (eMBB) communications to virtual reality, automated driving, Internet-of-Things, etc. 
Nevertheless, looking at future new uses of technologies, applications, and services, as well as at the recent predictions for the development of new technologies expected for 2030, it is already possible to foresee the need to move Beyond 5G (B5G) and to design new technological enablers for B5G connect-compute networks \cite{Calvanese2019}, incorporating new technologies to satisfy future needs at both individual and societal levels. While some near future technological solutions will be included in the long-term evolution of 5G, others will require a radical change, leading to the standardization of the new 6th Generation (6G).

This paper is a vision paper whose goal is to motivate a paradigm shift from the mainstream research, which basically builds on Shannon's framework, towards semantic and goal-oriented communications. In 1948, Shannon established the basis for a mathematical theory of communication, deriving the conditions ensuring the reliable transmission of a sequence of symbols over a noisy channel. In the following past 70 years, building on Shannon’s theory, the research on communications has produced a number of significant advancements, including multiple-input multiple-output (MIMO) communications, new waveform design, mitigation of multiuser interference in both uplink and downlink channels, etc. In parallel, a remarkable progress has been achieved in the network architecture, leading to  network traffic engineering, network function virtualization (NFV), software defined networking (SDN) and network slicing, which represent some of the key 5G network technologies. 

Today, while deploying the fifth generations (5G) of wireless communication systems and kicking-off research on beyond 5G (B5G) future networks \cite{Calvanese2019}, the need for a paradigm shift from Shannon's legacy begins to take shape. The motivation is dictated by the observation of the current trend witnessing the demand for wider and wider bandwidths to cope with the ever increasing request of higher data rates to accommodate for the incoming new services, like virtual reality or autonomous driving. Looking at already present trends, we can foresee, within a decade from now, scenarios where virtual and real worlds will be blended seamlessly. But the challenges arising from this never ending request are doomed to face, at some point, a bottleneck represented by the scarcity of resources, like spectrum and  energy. One example is the increase of carrier frequencies: As frequency increases, there is more room for wider bandwidth, but several undesired phenomena appear, like blocking, atmospheric absorption, power amplifier efficiency drop \cite{belot2020spectrum}, etc. 

The inspiration on how to handle the challenge represented by this never ending race comes again from Shannon and Weaver, who identified three levels of communication \cite{shannon1948mathematical}: (i) transmission of symbols (the technical problem); (ii) semantic exchange of transmitted symbols (the semantic problem); (iii) effect of semantic information exchange (the effectiveness problem). Shannon deliberately focused on the technical problem. At that time, this was indeed a very intelligent move, as it enabled him to derive a rigorous mathematical theory of communication based on probabilistic models. However, nowadays, the vision of the network as an enabler of pervasive intelligent services, with a strong emphasis on its effectiveness and sustainability, suggests that assuming semantics as irrelevant is no longer justified. Besides thinking about ``how'' to transmit, we need to focus on ``what'' to transmit. Or, quoting John von Neumann, {\it ``there's no sense in being precise when you don't even know what you're talking about''}. 

In this paper, starting precisely from the initial Shannon and Weaver's categorization, we propose a vision of 6G networks that incorporates semantics and effectiveness aspects, as recently suggested also in \cite{popovski2020semantic}.
The new vision takes inspiration from another visionary giant, Nikola Tesla, who stated, in 1926:  {\it ``When wireless is perfectly applied, the whole Earth will be converted into a huge brain''}. 
Following this idea, we believe that the network design can receive significant hints by observing how the human brain processes the signals perceived from the environment by its senses. The brain in fact learns from past actions (and from the culture accumulated by the humankind in its history), and takes complex decisions, in a short time, with a sustainable energy consumption. Mimicking this excellent example provided by nature, we believe that 6G networks should take semantics and effectiveness aspects as central aspects of network design. In this context, focusing on semantics and clearly identifying the goal of communication, helps us to distil the data that are strictly {\it relevant} to conveying the information intended by the source or to fulfilling a predefined goal. Disregarding irrelevant data becomes then a key strategy to significantly reduce the amount of data to be transmitted and recovered, thus saving in bandwidth, delay and energy.
According to this view, goal-oriented and semantic communications will be a keystone for exploring the meaning behind the bits and enabling brain-like cognition and effective task execution among distributed network nodes. This change of perspective represents a fundamental paradigm shift where success of task execution at destination (effectiveness problem) is the core concern rather than achieving error-free communications at symbol level (technical problem). In this context, information has no value unless it has an exploitable and {\it explainable} meaning \cite{xie2020explainable}. Explainability is an aspect that is gaining more and more importance because of the widespread use of deep neural networks (DNN). DNN's represent a very powerful computational tool that finds important applications in a variety of fields including, in particular, the design and effective control and management of new generation networks. But often it is not evident how an input of the DNN generates the corresponding output. More specifically, even if the relation between input and output is clear, the weights of the network are the result of a training phase involving a highly nonlinear optimization, whose final output does not come with theoretical guarantees.  In most cases, deep neural networks work pretty well, but it is not always clear why. A central tenet in {\it explainable machine learning} is that the algorithms must provide information allowing a user to relate characteristics of input features with its output. Going into an in-depth presentation of explainable deep neural networks goes beyond the scope of our paper. However, the interested reader can refer to \cite{xie2020explainable} for an in-depth review of the methods used to introduce explainability features in neural networks.

The paper is organized as follows. In Section \ref{6G Use Cases, KPIs and New Services}, we highlight the new use cases, report some important Key Performance Indicators (KPI) and illustrate some new services aimed to support the new use cases. In Section \ref{Beyond Shannon?}, we motivate the need to incorporate the so called semantic and effectiveness levels, in Shannon's sense, thus suggesting a new architecture that facilitates an efficient cross-layer design capitalizing on the new levels. In Sections \ref{Semantic Communications} and \ref{Goal-oriented communication}, we focus on semantic and goal-oriented communications, respectively. Section \ref{Learning-based design in a C4 framework} is then devoted to illustrate the importance of learning-based approaches in the design of the new networks, an aspect that is already assuming more and more importance in 5G networks as well. 
Finally, some conclusions are drawn in Section \ref{Conclusions}.


\section{6G Use Cases, KPIs and New Services}
\label{6G Use Cases, KPIs and New Services}
The design and engineering of a \textit{new generation} of wireless communication systems is motivated by the ambition to meet new societal challenges and to enable radically new use cases targeting new value creation. A \textit{new generation} builds upon the evolution of technologies already adopted and on a few new technological break-troughs and new network architectures that enable revolutionary new services. 
A fundamental question when starting research on the design of a new generation is whether the new generation should be backward compatible or clean slate. This is a never ending debate. Mechanics of economics and past experience encourage the view that \textit{the next generation} should be as much as possible backward compatible, to avoid huge CAPEX investments for renewing hardware in the network and terminals. The downside is that some possible breakthrough advantages that a totally revolutionary  technology might put forward might be either not supported by the new standard or not fully exploited.

From the one hand, the \textit{evolution toward Beyond-5G (B5G) networks} is shaped following the ``classical problem of wireless communications'', which is focused on achieving reliable and cost effective data communication over noisy channels. On the other hand, societal and environmental needs are stimulating radical changes into today economical approach to business and value creation. The societal acceptance of a new technology is already at a critical stage and future 6G networks are required to 
address societal and environmental issues rather than just creating new business opportunities and added value for operators, industry and IT companies. 
To this end, multiple technological enablers for beyond 5G networks are currently investigated following roadmaps to enable 6G services by 2030 \cite{Calvanese2019} \cite{Letaief20196GRoadMap} \cite{dore2020SubTHz}. Some of the use cases in 6G will evolve from the emerging 5G applications, others will rise from new societal and economical needs. 
In the following sections, we will first highlight some of the emergent new use cases, describe some of the relevant KPIs and then identify the new services aimed to provide an efficient deployment of the new use cases.

\subsection{6G Use Cases and KPIs}
%
%



Already today, we are experiencing how society and industry are becoming increasingly data-centric and automated. This phenomenon is expected to intensify in the next decade and beyond. The fusion of digital and real worlds and the support of networked intelligence and automation are driving the next technological revolution. The boundary between computer science, artificial intelligence and telecommunications is disappearing, creating the momentum for a plethora of new applications and challenging the future 6G networks with the ongoing race between cost and complexity of delivering new services. 
The ITU 2030 group published a first speculative vision on future 6G services and use cases \cite{network203020196Gusecases}, identifying the evolution of virtual reality (VR) and mixed reality (MR) services as a main driver for future 6G services. A list of possible new use cases motivating the move toward 6G networks is listed in Table \ref{tab:6G-Use Cases}.




\begin{table}[ht!]
\label{tab:6G-Use Cases}
\centering
\begin{tabular}{ | l | c | c | } 
\hline
\textbf{Use Case} & \textbf{Service}  \\ 
\hline 
\hline Internet of nano-bio things \cite{Zhang2019UseCase6G} & 1\\
\hline Connected living & 1,2\\
\hline Precision and personalized digital health \cite{Yang2019UseCase6G} &  1,3  \\ 
\hline Remote areas connectivity \cite{Nandana2020-6GBroadbandWhitepaper} & 2  \\
\hline Space connectivity \& 3D AI support \cite{Calvanese2020Sky6G} & 2 \\
\hline Smart railway mobility  \cite{Nandana2020-6GBroadbandWhitepaper} & 2   \\
\hline Extreme capacity Xhaul \cite{Nandana2020-6GBroadbandWhitepaper} & 2  \\ 
\hline Precision agriculture \cite{Himesh2018SmartAgriculture} & 1,2,3\\
\hline Unmanned mobility & 1,2,3\\ 
\hline Autonomous vehicle mobility \cite{Nandana2020-6GBroadbandWhitepaper} & 1,2,3 \\
\hline Integrated super smart cities \cite{Tariq2020UseCase6G}& 1,2,3\\ 
\hline Space travel \cite{Zhang2019UseCase6G} & 1,2,3\\
\hline Multi-sense Services  \cite{Ericsson6GSenses} & 3,4\\
\hline Multi-sensory and mobile immersive eXtended reality (XR) \cite{Nandana2020-6GBroadbandWhitepaper} & 3,4 \\ 
\hline Multi-sensory holographic teleportation & 3,4 \\
\hline Multi-sensory haptic communications for Virtual and Augmented Reality & 3,4 \\
\hline Multi-sensory affective computing & 3,4\\
\hline Holographic communication \cite{david20186g}  \& telepresence \cite{Calvanese2019} &  3,4  \\ 
\hline Remote XR education & 3,4\\
\hline Consumption of digital experiences over physical products   & 3,4 \\
\hline Human centric AI support & 3,4\\
\hline Brain-to-computer interactions \cite{Calvanese2019} \cite{saad2019vision} & 3, 4\\
\hline Wireless robotics \cite{saad2019vision} &  1,3,4 \\ 
\hline Sustainable connectivity \& AI support & 1,2,3,4\\
\hline Society 5.0 \cite{shiroishi2018society} & 1,2,3,4\\
\hline High precision manufacturing, remove monitoring \& control \cite{Calvanese2019}&  1,2,3,4 \\ 
\hline Intelligent Disaster Prediction \cite{Yang2019UseCase6G} & 1,2,3,4\\
\hline Smart Digital-Twin Environments & 1,2,3,4\\ 
\hline Bidirectional intelligence intertwining (natural and/or artificial) & 4\\
\hline
\end{tabular}
\caption{6G Use cases and associated services: 1 MMTCCxDI, 2 GeMBB, 3 URLLCCC, 4 Semantic.}
\end{table}

While the vision on what 6G should be is still evolving, academia, industry and standardization bodies are already working on identifying candidate KPIs for future 6G services, use cases \cite{Calvanese2019} and applications \cite{Giordani2020} \cite{nakamura20205g}. Some tentative numbers are reported in Table \ref{tab:6G-KPIs}.

\begin{table}[ht!]
\label{tab:6G-KPIs}
\centering
\begin{tabular}{ | l | c | c | } 
\hline
KPI& \textbf{5G} & \textbf{6G} \\ 
\hline \hline
Traffic Capacity & 10 Mbps/$m^2$ &  $\sim$ 1-10 Gbps/$m^3$ \\ 
\hline Data rate DL & 20 Gbps &  1 Tbps \\ 
\hline
Data rate UL & 10 Gbps &  1 Tbps \\ 
\hline
Uniform user experience & 50 Mbps 2D everywhere &  10 Gbps 3D everywhere\\ 
\hline
Mobility  & 500 Km/h & 1000 Km/h \\
\hline
Latency (radio interface) &  1 msec &  0.1 msec \\ 
\hline
Jitter & NS &  1 $\mu$sec \\
\hline
Communication reliability & $1-10^{-5}$ & $1-10^{-9}$\\
\hline
Inference reliability & NS & TBD\\
\hline
Energy/bit & NS &  1 pJ/bit\\
\hline
Energy/goal & NS &  TBD\\
\hline
Localization precision & 10 cm on 2D &  1 cm on 3D\\
\hline
\end{tabular}
\caption{Comparison of 5G and 6G KPIs being discussed; NS= Not Specified: TBD= To Be Defined case-by-case.}
\end{table}

There will be two families of 6G KPIs: i) the set of KPIs already envisaged in 5G networks, such as  peak data rate, area traffic capacity, connections density, communication reliability, end-to-end latency, spectrum efficiency, energy efficiency, etc., marking the conceptual continuity with 5G; ii) an entirely new class of KPIs that will support the vision toward a computation-oriented communication network, where distributed artificial intelligence will play a prominent role. The new KPIs will need to take into account, for example,  the reliability of the decisions taken by intelligent agents present in the network, the time needed to take decisions,  etc.

A broad class of new use cases will need to support extended reality (XR) applications, realizing an entire {\it reality-virtuality continuum}, where both reality and virtuality may be augmented. Augmented reality (AR) will support applications where human perception is augmented by computer-generated information. Conversely, virtual reality reality replaces the real-world environment with a simulated one. Within such a context, augmented virtuality (AV) includes applications where a real world object is inserted into a computer-generated environment \cite{maier2020toward}. This may be helpful, for example, for an engineer checking the introduction of a new manufacturing tool within a smart factory controlled, from remote, through a VR application. 

\textit{Holographic} communications \cite{david20186g} and \textit{multi-sense} \cite{Ericsson6GSenses}, including haptic, communication services are also expected to take place in a not-so-distant future. Such applications will dominate services not only in the realm of entertainment, teleconferencing, smart working and tourism, but they will also enable more life-impacting and industrial productivity applications. Those futuristic services will be for instance key to the Japanese {\it Society 5.0} vision \cite{shiroishi2018society} or for remote holographic presence \cite{Tariq2020}, industrial maintenance in hostile operational environments and intelligent production at broad. Such family of new use cases will impose stringent requirements in terms of per-link capacity. For instance, holographic communications employing multiple-view cameras are expected to require several terabits per second (Tb/s) per link in both uplink and downlink \cite{li2018towards} \cite{berardinelli2018beyond} (a requirement not supported by 5G) and a stringent end-to-end (E2E) latency to ensure real-enough virtual and seamless remote experience. 6G targets also services, such as industrial automation, autonomous system, and massive networks of sensors, in which machines, and not humans, are the endpoints. These communicate-and-compute services will require new stringent requirements in terms of latency {\it and its jitter}, in order to ensure a seemingly deterministic performance of the network  \cite{calvanese2020IIoT}. Furthermore, extremely high reliability will be required to improve performance not only at the physical to networking layer but also on inference-based intelligent mechanisms supporting them. Clearly, the specific targets on both communication \cite{berardinelli2018beyond} and inference reliability will depend on the specific use case. With 6G, new applications will not be limited to the realm of entertainment and teleconferencing, but more disruptive applications begin to emerge, some of which are life-impacting while others provide alternative solutions for intelligent production and smart mobility with multi-dimensional transportation network consisting of all ground-sea-air-space vehicles with peak mobility up to 1000 Km/h (see, for instance, the hyperloop transportation system \cite{vora2020provisioning}). This multi-dimension mobility vision opens also the opportunity for Three-Dimensional \textit{(3D) native services}, enabling end users and machines moving  in the 3D space to perceive seamless 6G service support and teleport cloud functionalities on demand, {\it where and when} the intelligence support is needed in the 3D space \cite{Calvanese2020Sky6G}. To this end, KPIs such as localization precision and uniform user experience will be defined in both Two-Dimensional (2D) and 3D space.\\


Of course, not all the KPIs shown in Table \ref{tab:6G-KPIs} will have to be achieved simultaneously, all the time, everywhere, in every possible condition. Only a selected subset of KPIs should be attained locally in space and time,
depending on future 6G application and service needs, with a high degree of flexibility and adaptivity.
The targeted 6G performance improvement in terms of data rate, latency at radio interface and network energy efficiency, etc. follows the well established performance driven KPIs mechanics: a technology intense evolution across generations. This translates in imposing a factor 10 to 100 of KPIs' improvement between successive wireless networks generations. The rational is to deal with the expected exponential traffic growth \cite{itu2015expraffic} and more immersive and interactive foreseen services. The most representative trend in today wireless communications is the unrelenting increase in signal bandwidths to achieve higher link capacity, increasing the network capacity and improve the user's QoE. This is also a major well accepted \textit{mega-trend} for 6G: for example, one goal is to achieve a factor of 100 increase in capacity by accessing the huge available bandwidth either in the sub-THz D-band (above 90 GHz) or in the visible light spectrum  \cite{Calvanese2019}, \cite{dore2020SubTHz}, \cite{haas2020Lifi6G}.\\

Up to now, generations of wireless systems have been designed to accomodate the exponential growth of downlink traffic. Nevertheless, starting from 4G, we experience a reduction and sometime inversion of the asymmetry between uplink and downlink traffic \cite{oueis2016uplink}.
Even though there are no precise forecasts on uplink traffic evolution, the traffic pattern change is inevitable. The uplink traffic is exploding at much faster rate than the downlink traffic. Already with 4G networks, a study by Nokia Simens Netowrks showed how the overall usage ratio between uplink and downlink reached already approximately 1:2.5 \cite{solutions2013networks} for services like pear-to-pear TV, pear-to-pear sharing, massive IoT and cloud support. This is due to the introduction of the cloud support and the rising use of content sharing platfoms. A larger share of data is crossing the network, conveyed on the uplink between a huge number of connected devices, collecting large amounts of data and requiring a pervasive support of offloading services to the cloud (computation and storage). In 5G, the increasing support of machine learning algorithms is causing a further explosion of the uplink traffic. 5G uplink capacity  has not been sized to meet such exploding demand for the next decade. To accommodate for the rising share of uplink traffic, similar capacity requirements are foreseen for uplink and downlink channels in 6G. In addition, device-to-device (D2D) communication will consume an increasing share of the network capacity, defining a novel layer of communication.\\

A major leap kicked off with 5G is the increasing interplay between communications and computation.
With 6G, this trend will be further intensified through the introduction of an increasingly number of distributed intelligent nodes collecting, processing, and storing data. Some identified use cases, such as Industrial IoT or virtual reality, already impose new KPIs requirements such as stringent latency bounds, packet delivery jitter, reliability or achievable throughput, but also regarding the systems dependability, i.e. the ability to make guarantees for a network {\it deterministic} behavior \cite{calvanese2020IIoT}. An example of application is remotely controlled high-precision manufacturing, requiring jitter delays in the order of a microsecond \cite{berardinelli2018beyond}. These requirements are typically quite distinct from those that have traditionally guided the design and deployment of public 5G networks.\\ 


Nevertheless, in our view, the most remarkable  feature of 6G will be not only the performance improvement in terms of typical KPIs. 6G will need to produce a paradigm shift reflecting a {\it degrowth-by-design} principle, leading to the introduction of a new class of KPIs. 
The new perspective reshapes the network as a truly pervasive computing system enabling new interactions among humans and machines and intelligent services with {\it sustainable} costs, in economical and ecological terms. 5G already represents a significant step forward in  this direction; 6G will take this perspective as its driving principle. Within this perspective, new KPIs will come into play, such as learning reliability or energy consumption associated to goal accomplishment, depending on the specific services.
Furthermore, with 6G, energy-related KPIs will have to deal not only with network energy  consumption \cite{earth2009} or terminal battery life extension. The ambition is to achieve, wherever possible, battery-free communications, targeting, in some applications, communication efficiency on the order of 1 pJ/b \cite{IEEEVTM-Liu2016}. Moreover, since 6G operation will be intensively supported by machine learning and artificial intelligence, specific energy constraints will be defined across the whole generation-to-processing data chain.\\


\subsection{6G New Services}
6G services are foreseen to be operational starting from 2030, and for the next ten to five-teen years. Some of those services will be first offered with 5G technologies, or its long term evolution; others will require disruptive technologies and completely new network operations to meet their stringent requirements, following the usual never-ending technology growth model. 
%
In our view, besides the services already supported by 5G networks, to accommodate for the plethora of new use cases, 6G will incorporate the following  services:

\begin{itemize}
\item  \textbf{\textit{Massive Machine Type Communications supporting Distributed Intelligence}} (MMTCxDI) services
- Following the paradigm shift initiated with 5G, future 6G Machine-Type Communications (MTC) will enlarge the capabilities of massive MTC already foreseen in 5G to include pervasive distributed computation algorithms supporting the distributed intelligence of 6G networks. Criticality, effectiveness and scalability will be intrinsic features of this new service. Typical applications will include intelligent transportation systems, connected living, super smart cities, etc.
\item  \textbf{\textit{Globally-enhanced Mobile BroadBand}} (GeMBB) services
- This service will expand the computation-oriented communication environment to support rate-hungry applications, such as extended reality services, also in remote locations, such as rural areas \cite{Yaacoub6GRuralAreas}, oceans, and the sky \cite{Calvanese2020Sky6G}, {\it on demand}, i.e. when and where needed. 
\item  \textbf{\textit{Ultra-Reliable, Low Latency Computation, Communication and Control}} (URLLCCC)
- This service will extend the capabilities of URLLC services already supported by 5G networks, in order to incorporate the computation services running at the edge of the network and E2E (remote or automated) control. In this new service, reliability and latency refer not only to the communication aspect, but also to the computation side, like, e.g., learning accuracy or correct classification probability. The use cases supported by this new service will include factory automation,  multi-sensory XR \cite{saad2019vision}, connected and autonomous terrestrial and flying vehicles, etc.

\item \textbf{\textit{Semantic Services}}
%
- These services will support all applications involving a share of knowledge between the interacting parties. The applications will not be bounded to human to human (H2H) interactions, but will involve also human to machine (H2M) and machine to machine interactions (M2M). Goal of this service will be the seamless connection and intertwining of different kinds of intelligence, both natural and artificial. Empathic and haptic communications, affective computing, autonomous and bi-directional collaboration between different Cyber-Physical-Spaces (CPS), etc., will be supported. This new type of service will offer \textit{intelligence as a service}, bringing a radical paradigm shift that will revolutionize wireless services from connected things to connected intelligences. 
\end{itemize}
Although related to  the \textit{Semantic Web} paradigm \cite{Berners2001SemanticWeb}, semantic communications differs substantially from the semantic web. The semantic web is that ``version'' of the web that associates to each document (a file, an image, a text, etc.) information and metadata which, by providing a semantic context, facilitates the automatic query and interpretation by a search engine. 
Certainly, semantic communications will benefit from some of the technologies developed for the semantic web, such as ontologies for example, but the scope of semantic communication is to set-up a more efficient communication system, exploiting the knowledge shared a priori between transmitter and receiver, such as a shared language or shared models.

\section{Beyond Shannon? A new architecture}
\label{Beyond Shannon?}
Until the 5G era, communication has been the basic commodity of every wireless generation. The key challenge has been the reduction of the uncertainty associated to the correct reception of exchanged data, while targeting higher capacity and reliability and lower latency. Such legacy of Shannon's model has pushed a never-stopping race for broader bandwidths, thus exploring higher frequency bands.
Since the deployment of 4G, the energy consumption of network and wireless devices has limited the practical services' operation, pushing research to approach the theoretical communication limits, established by Shannon, while optimizing the use of available resources.

With 5G, the communication network has evolved towards a communicate-and-compute system, where the support of the (edge) cloud has fed the cybernetic vision of Norbert Wiener, where communicate-compute-control tasks generate \textcolor{black}{a continuous loop involving sensing, computing, controlling and actuating, laying the foundations for the birth of intelligent machines}. 6G services will induce a further drastic change on the conventional notions of knowing and learning, guessing and discovering. This will require significant advances on the communicate-and-compute infrastructure, paving the way to making knowledge and decision a commodity of next generation networks. In such a futuristic context, information accumulates at a rate faster than what can be filtered, transmitted and processed by some kind of intelligence, either natural or artificial. 

Keeping in mind the inevitable limitedness of available resources, the challenge is to design the new network, while respecting a {\it degrowth} principle. 
The key question is: Can we deliver more intelligent mobile services without necessarily requesting for more capacity, more infrastructure (communication, computation, storage), more energy?
We believe that this challenge cannot be handled efficiently only resorting to higher data rates, possibly exploiting wider bandwidths. \textcolor{black}{Most likely, 6G will still rely on millimeter-wave technologies and will complement them with sub-Terahertz and Visible Light Communications. But our vision is to make the communication networks {\it qualitatively} more efficient, without necessarily running an endless race towards increasing resources, but rather inventing a more intelligent use of them.} The hint on how to proceed comes again from Shannon's ideas.
In their seminal work, Shannon and Weaver suggested that the broad subject of communication can be organized into three levels  \cite{weaver1953recent}:
{\it 
\begin{description}
    \item [{\it \textbf{Level A}.}] How accurately can the symbols of communication be
transmitted? (The technical problem.)
\item  [{\it \textbf{Level B}.}] How precisely do the transmitted symbols convey the
desired meaning? (The semantic problem.)
\item [{\it \textbf{Level C}.}] How effectively does the received meaning affect 
conduct in the desired way? (The effectiveness problem.)
\end{description}
}
Shannon \cite{shannon1948mathematical} provided a rigorous and formal solution to the {\it technical problem}, lying the foundations of what is known today as information theory. 
Shannon left deliberately aside all aspects related to semantic and effectiveness. However, now that communication is becoming a commodity enabling a variety of new services, interconnecting humans and machines possessing various degrees of intelligence (either natural or artificial), the semantic and effectiveness aspects become preeminent actors that can no longer be left aside.
In our view, 6G networks will have to evolve to incorporate all the three levels foreseen by Shannon and Weaver, as also recently suggested in \cite{popovski2020semantic}.

\begin{figure}[ht]
    \centering
    \includegraphics[width=\columnwidth]{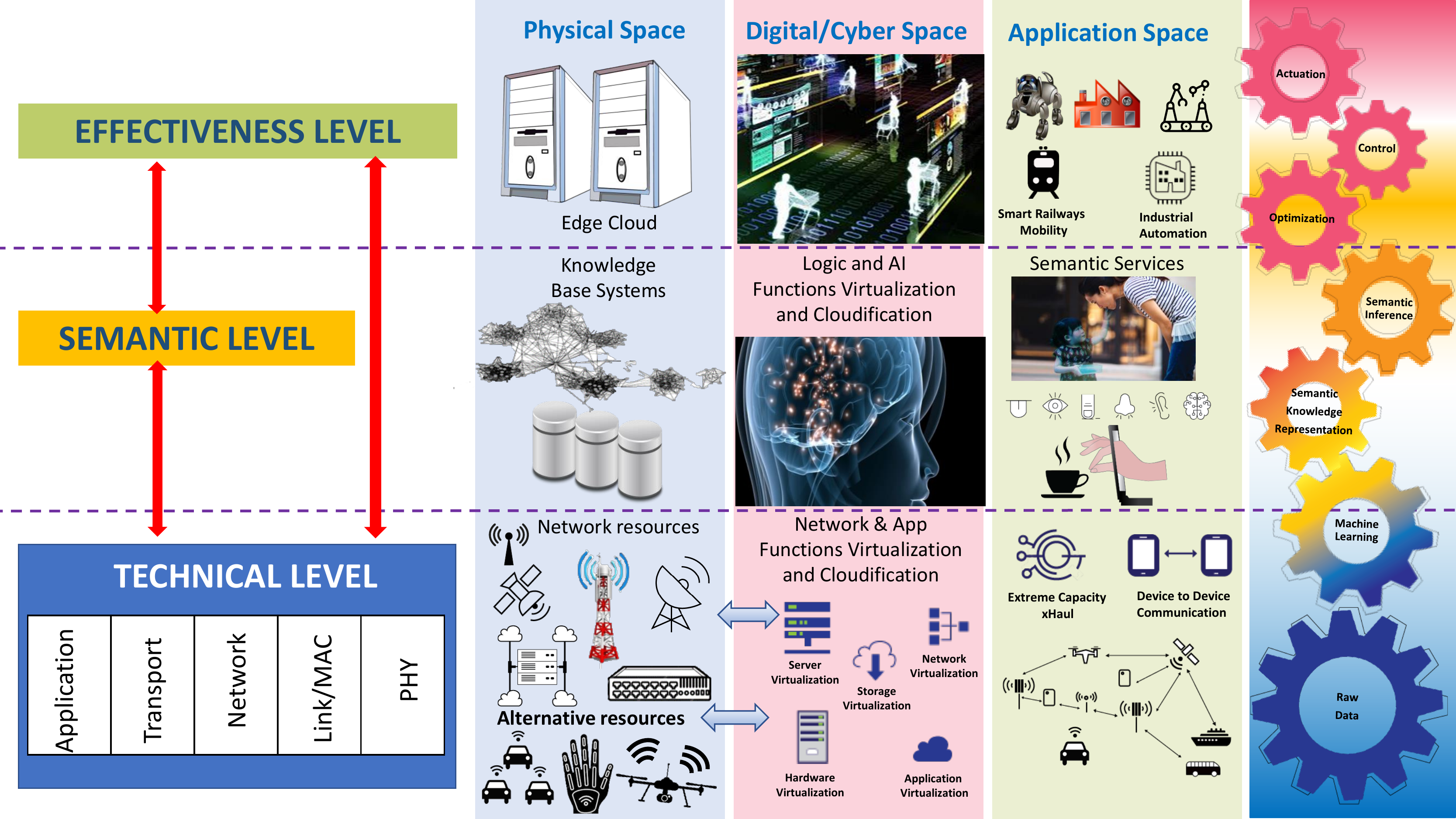}
    \begin{center}
        \caption{Interaction of the physical-semantic-effectiveness three-levels architecture and communication protocol stack (left) with physical, digital and application spaces (right).}
    \end{center}
    \label{architecture}
\end{figure}

The proposed new three-levels architecture for 6G is represented schematically in Fig. \ref{architecture}.
Fig. \ref{architecture} has five columns. From left to right: the three-levels architecture and communication protocol stack, the seamless blending of physical, digital and application spaces, and the machinery enabling this view, based on artificial intelligence tools, including knowledge representation and machine learning, running on the data collected across the network, and closing the loop with the control and actuation functionalities. \\
On the left side, the new protocol stack includes the three levels foreseen by Shannon and Weaver. At the bottom level, there is the \textit{technical level} incorporating the typical protocol stack of nowadays communication networks. Taking advantage of the 5G experience, the technical level builds heavily on virtualization techniques, NFV and SDN. However, besides the virtualization of network functionalities, the new computation-oriented communication network includes now also the virtualization of many application-layer functionalities. Virtualization is expected to play a key role in 6G, even bigger than in 5G, as it will include not only network functionalities, but also functions associated to the semantic and effectiveness levels. This additional virtualization is fundamental to distribute computation and communication tasks across densely distributed  virtual machines (or containers) properly orchestrated from the semantic and effectiveness level. 

On top of the technical level, there is the \textit{semantic level}, at least for all services where semantics has a well defined role. On top of the semantic level, there is the \textit{effectiveness level} that orchestrates the lower levels in order to optimize the use of resources and meet the service KPIs. Since not all use cases include a semantic aspect, the effectiveness level is also  allowed to interact directly with the technical level. As will be explored in the next sections, the semantic and effectiveness levels allow also the proper cross-layer interaction between the layers composing the technical level. This is indeed a key aspect required for an intelligent allocation of network and computation resources.\\

We briefly illustrate now the main research directions motivating the need to move beyond Shannon. 
\begin{itemize}
    \item \textit{Semantic communications}: Communication among humans involves the exchange of information, where the word information is associated to meaning; in conveying a concept from source to destination, the relevant aspect is {\it what} is communicated, i.e. the information content, not {\it how} the message is brought to the destination. A correct semantic communication occurs if the concept associated to the message sent by the source is correctly interpreted at the destination, which does not necessarily imply that the whole sequence of bits used to transmit the message be decoded without errors. Intuitively speaking, one of the key reasons why the semantic level offers a significant performance improvement with respect to the pure technical level is because it exploits the share of a priori knowledge between source and destination. This knowledge may be the human language or, at a more general level, a formal language consisting of entities and a set of logical rules that allow the receiver to correct errors occurring at the symbolic level.  An interesting aspect of semantic communication is the interplay between different languages and the intertwining of natural  and artificial intelligence.
    \item \textit{Goal-oriented communications}:
    Communication among interacting entities is often carried out to enable the involved entities to accomplish a joint goal. The fundamental system specification is then associated to the goal, its correct accomplishment, within a given time constraint, using a given amount of resources (energy, computation, etc.). The communication system enabling the interactions among the entities involved in the goal should be then defined in order to focus on the goal-related specifications and constraints. This means for example, that all information not strictly relevant to the fulfillment of the goal can be neglected. The effectiveness level is the level responsible for the efficient management of goal-oriented communications. It will exploit semantic aspects, whenever appropriate and relevant, and it will act by properly orchestrating the resources available at the technical level, including the network nodes, the computation, control and actuation devices.
    \item \textit{Online learning-based communication and control}: The increasing pervasive introduction of machine learning tools in all the layers of the computation-oriented network yields a further breakthrough in the network design. On the one hand, online machine learning algorithms provide the possibility to reshape traffic, change coding and decoding strategies, scheduling, etc., as a function of an online monitoring of the network, thus enabling an efficient use of resources. On the other hand, the communication network enables the capillary distribution of machine learning tools, to accommodate for stringent delay constraints. This overall scenario calls for a joint orchestration of computation, communication, storage and control resources.  
\end{itemize}

In the following sections, we will dig into the  fundamental challenges and opportunities associated to the above topics. 

\section{Semantic Communications}
\label{Semantic Communications}
The goal of this section is to explore the potentials offered by the introduction of the semantic level, as shown in Fig. \ref{architecture}.
Several schools of thought have proposed different alternative approaches to generalize Shannon's information theory, each aimed at emphasizing different perspectives: philosophy of information \cite{floridi2002philosophy}, logic and information \cite{devlin1995logic}, information algebra \cite{kohlas2012information}, information flow \cite{barwise1997information}, quantum information theory \cite{nielsen2002quantum}, algorithmic information theory \cite{calude2013information},  \cite{chaitin1977algorithmic}.\\
Building on a more general information theory, there are also various proposals concerning the design of a semantic communication system
\cite{willems2005semantic}, 
\cite{juba2011universal}, \cite{bao2011towards},
\cite{goldreich2012theory}, \cite{basu2014preserving}, 
\cite{guler2018semantic}, \cite{kountouris2020semantics}. In \cite{willems2005semantic}, it was proposed a method to perform semantic lossless data compression, as a way to produce a significant compression with respect to entropy-based encoders. 
Semantic data compression and the capacity of a semantic channel have been studied in \cite{bao2011towards}.
An end-to-end (E2E) semantic communication framework incorporating  semantic inference and physical layer communication problems has been proposed and analyzed in \cite{guler2018semantic}, where semantic was exploited by considering similarities between single words. A further significant extension was given in the recent work \cite{xie2020deep}, where the authors, building on recent natural language processing (NLP) tools, use a deep neural network to learn, jointly, a semantic/channel encoder, considering similarities between whole sentences. The key aspect of  \cite{xie2020deep} is to recover the meaning of the transmitted message, rather than avoiding bit- or
symbol-errors, as in conventional communications.
 
Before delving into the technical problems associated to the definition of a semantic communication system, it is necessary to clarify what do we mean by semantics. In linguistics, semantics is the study of meaning in human languages. On its turn, meaning is a relationship between two sorts of things: signs and the kinds of things they intend, express, or signify. A human language is the ensemble of signs associated to things in real world or to abstract thoughts and the rules used to compose these signs. Each language has a {\it structure}, given by the set of rules used to compose its signs to create sentences that are meaningful. This definition of meaning and language can be extended to artificial languages, like for example a computer programming language, after proper identification of symbols and rules.

A genuine theory of semantic information should be a theory about the information content, or meaning, of a message, rather than a theory about the symbols used to encode the messages. To distinguish between the different interpretation of the word information, in the following we will use the term {\it semantic information}, to refer to information  as associated to a meaning, and the term {\it syntactic information}, in the Shannon's sense, which is associated to the probabilistic model of the symbols used to encode information. 

As a very simple example, pressing the keys of a computer keyboard at random generates a message that has a high syntactic information, because the generated symbols are approximately independent and uniformly distributed, so that their entropy (average  information in Shannon's sense) is maximum. However,  most likely, the generated message carries zero semantic information, as it does not carry any meaningful content. 
\subsection{Semantics and knowledge representation systems}
Semantic information is associated to the level of knowledge available at the source and destination sides. In general terms, quoting Dretske \cite{dretske1981knowledge}, ``{\it information is that commodity capable of yielding knowledge, and what information a signal carries is what we can learn from it.}'' From this perspective, a semantic communication from source to destination occurs correctly, or with a high degree of fidelity, under the following circumstances:
\begin{enumerate}
\item the destination is able to recover from the received message a content (meaning) that is equivalent to that of the message emitted by the source;
    \item the destination is able to increase its level of knowledge thanks to the received message.
\end{enumerate}
Semantic equivalence means that the meaning intended by the source of the message is {\it equivalent} to the meaning understood by its destination. There might be several sets of symbols, which convey the same meaning, even though they have a completely different structure.  This way of looking at information marks a significant departure with respect to the way information is used in Shannon's information theory, in at least three respects: 
\begin{enumerate}
    \item the amount of information conveyed by a message is associated to its semantic content, and it is not necessarily related to the probability with which the symbols used to encode the message are generated;  
    \item in semantic communication, what matters is the specific content of {\it each} message, and not the average information associated to all possible messages that can be emitted by a source;
    \item the amount of information conveyed by a message depends not only on the message itself, but also on the level of knowledge available at source and destination, at the time of communication.
\end{enumerate}  
Since semantics (meaning) is associated to a knowledge system,  dealing with the semantic of a message requires first a formal way to represent knowledge. Knowledge Representation (KR) and reasoning is indeed one of the cornerstones of artificial intelligence \cite{russell2010artificial}. The goal of KR is the study of computational models to represent knowledge by symbols and by defining the relations between symbols, in a way that makes possible the production of new knowledge. Among the many alternative ways to represent knowledge, {\it graph-based} knowledge representation plays a key role \cite{chein2008graph}. An example of graph-based representation is given by a {\it conceptual graph}, whose nodes are associated to entities, whereas the edges represent relations among entities \cite{chein2008graph}.  Given the vastness of knowledge, it is unthinkable to represent all knowledge within a single framework. The only viable approach is to build knowledge base (KB) systems associated to specific {\it application domains}. For each application domain, a KB is typically composed by a computational ontology, facts, rules and constraints \cite{chein2008graph}. A computational ontology provides a symbolic representation of the objects belonging to the application domain, together with their properties and their relations. Furthermore, besides the ontology, a KB system includes a {\it reasoning engine}, built on the rules and the constraints associated to the given application domain. The goal of the reasoning engine is to answer questions posed within the application domain.\\ 

A key aspect of a KB system worth pointing out is that a KB cannot be assumed to be able to provide a complete picture of the application domain it refers to. This happens because, even in a restricted domain, each object might have relations with a huge number of other facts or objects, so that it would not be possible to encompass all these relations. As a consequence, the {\it incompleteness} of the description is a central feature of a knowledge-based system \cite{chein2008graph}, and it represents a key distinction with respect to a database. Furthermore, incompleteness of a KB comes also from computational constraints as a complete reasoning might be very time-consuming. As a consequence of its incompleteness, a KB system might not be able to answer, for example, to the question if a statement is true or false, within a given time interval. Conversely, the need to provide an answer, while respecting a time constraint, typically results in an answer that is correct but only within a certain degree of reliability.\\

It is worth to point out that, in general, the KB available at the source, say ${\rm KB}_S$, may differ from the KB available at destination, say ${\rm KB}_D$. We say that a message is correctly interpreted at the destination node, according to ${\rm KB}_D$, if its interpretation is equivalent to that given at the source node, according to ${\rm KB}_S$, or if it induces a valuable modification of the destination KB, either in its ontology or in the definition of the reasoning rules. In the case of graph-based KRs, the change of the KB is reflected into a change of the graph. This change becomes then a possible way to measure the increase of knowledge carried by a message.\\

A key feature of a KB system is that the inference made on a message should depend only on the semantic, i.e. meaning, of the message and not on its syntactical form. This means that there could be alternative ways to encode the same concept into formally different sequences of symbols, all of which should give rise to the same semantic representation. As a simple example, the answer to the question ``how much is two plus two'' could be the sound ``four'' or it could be the symbol $4$ written on a piece of paper. The encoding mechanism and the number of bits necessary to encode the two messages would be totally different, but the semantic information would be exactly the same.
\subsection{Semantic source and channel coding}
\begin{figure}[ht]
\centering
\includegraphics[width=\columnwidth]{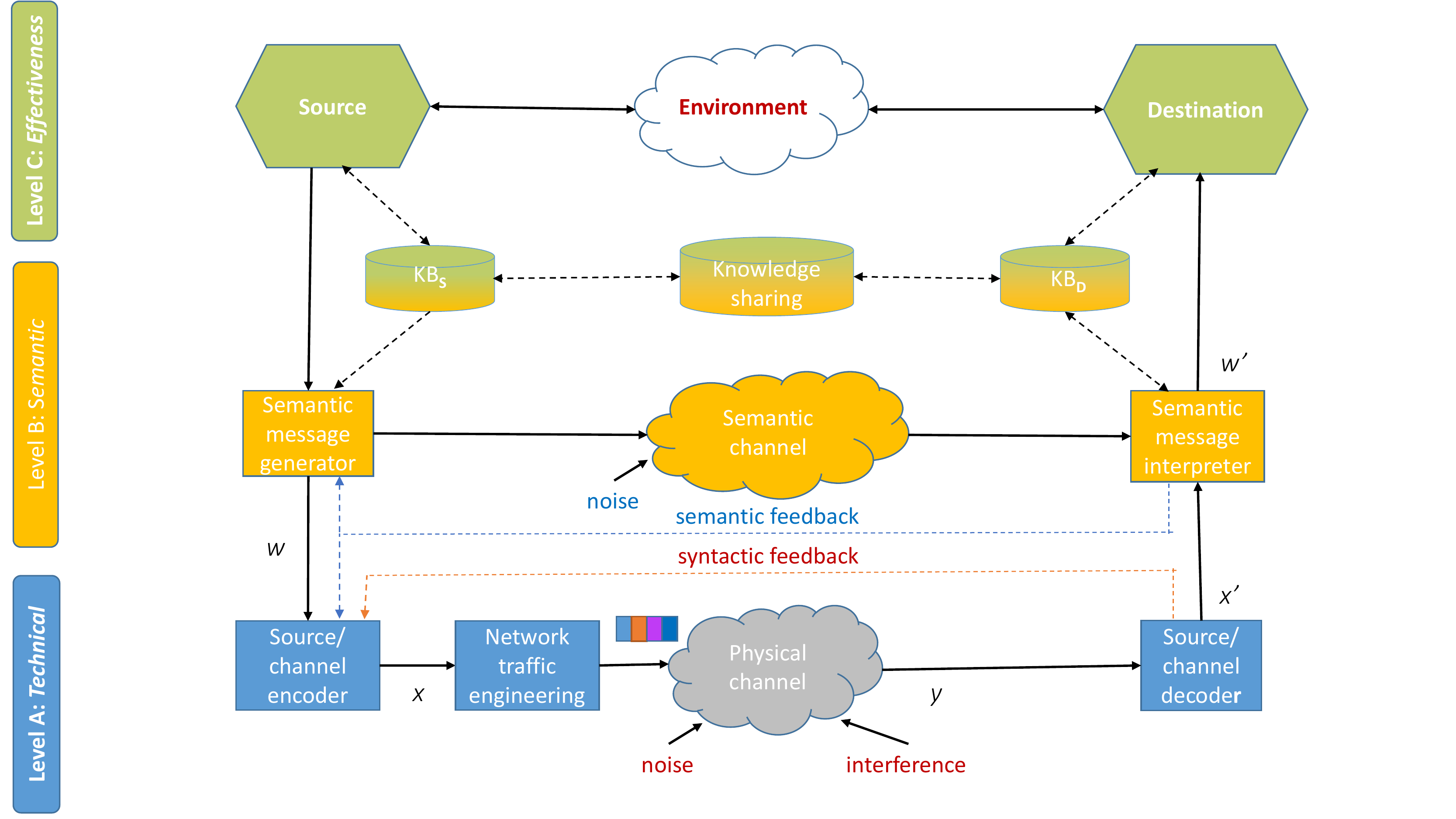}
\caption{Multi-level communication system.}
\label{Semantic-comm}
\end{figure}

A communication system incorporating the three levels of communication mentioned by Shannon and Weaver can be represented as in Fig. \ref{Semantic-comm}.
The block diagram shown in Fig. \ref{Semantic-comm} depicts three layers, as associated to the three levels of communication: technical, semantic and effectiveness. At the effectiveness level, there are two entities, a source $S$ and a destination $D$, that interact with each other through an environment. Source and destination nodes could be humans, machines or, in rather general terms, {\it agents}, where, in AI terminology, an agent is  something that can ``operate autonomously, perceive the environment, persist over a prolonged time period, adapt to changes, create and pursue goals'' \cite{russell2010artificial}. In particular, we consider {\it rational agents}, i.e. agents that act so as to achieve the best outcome of their acts. An agent could be a human, a machine, or a software. The scope of the interaction can be very broad in nature: sensing, controlling, extracting information from the environment, exchanging information, etc. To interact, the source $S$ generates a message $m \in {\cal M}_s$, belonging to a source alphabet ${\cal M}_s$, conveying the  semantic information that $S$ wishes to share with $D$. This message $m$ is generated according to the ontology and the rules given by the knowledge system ${\rm KB}_S$ available at the source. For instance, a concept could be represented, equivalently, by a speech signal or a by a text, produced using a given language. 
\begin{figure}[ht]
{\centering
\includegraphics[width=12cm]{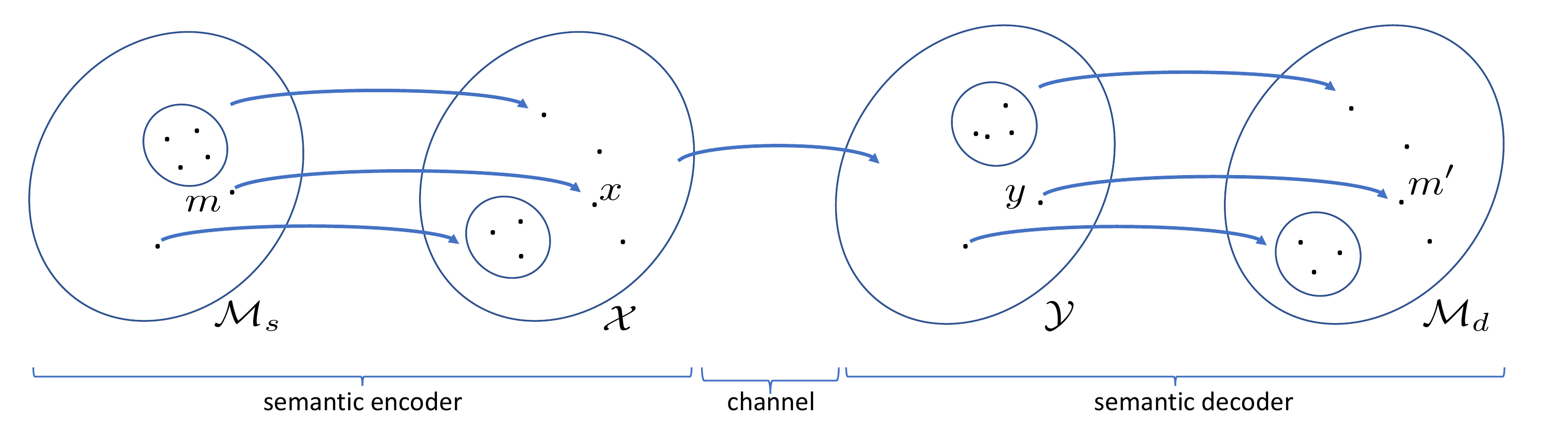}
}
\caption{Message-to-symbol mapping.}
\label{message mapping}
\end{figure}
To be physically conveyed to destination through a physical medium, the message $m$ is first translated into a sequence  $x \in {\cal X}$ of symbols, where ${\cal X}$ represents the symbols' alphabet, and then transformed into a physical signal suitable for propagation through the channel.

As depicted in Fig. \ref{message mapping}, the mapping from ${\cal M}_s$ to ${\cal X}$, denoted as $x=f(m)$, is not always one-to-one. Sometimes, the mapping is one-to-many. This happens when a message can be represented by multiple symbols, all conveying the same meaning. As an example, the symbol ``4'' is semantically equivalent to the English word ``four'' or to the speech signal of a speaker pronouncing the same word in English.
Sometimes, the mapping is many-to-one. This is the ambiguity problem that natural languages suffer from. For example, literally speaking, ``putting money in the bank'' could mean depositing money in a financial institution or burying it by the riverside. Usually, it is the context that helps to solve the ambiguity. Not surprisingly, disambiguation, dealing with teaching machines to solve the ambiguities of natural languages, is one of the key challenges of NLP.

According to Shannon's information theory, the translation from $m$ to $x$ typically includes a source encoder, to reduce the redundancy contained in the message, followed by a channel encoder, introducing structured redundancy to increase the communication reliability. Rules and properties of source and channel encoders follow the principles of Shannon information theory. The combination of source and channel encoder is denoted as a {\it syntactic} encoder, as it affects only the {\it form} of the message, but not its semantic content.
The sequence $x$ is then transformed into a physical signal, like an electromagnetic wave or an acoustic wave, to make it well suited to pass through the physical channel available for communication. 

Let us consider now a {\it semantic} encoder. Let us assume that the source is random and emits messages with probability $p_{{\cal M}_s}(m)$, $m \in {\cal M}_s$. The source has then a {\it message} entropy (in Shannon's sense)
\begin{equation}
\label{message-entropy}
    H_{\cal S}(M)=- \sum_{m\in{\cal M}_s}p_{{\cal M}_s}(m)\, \log_2 p_{{\cal M}_s}(m).
\end{equation}Proceeding as in \cite{bao2011towards}, denoting by $p_{{\cal M}_s}(m)$ the probability that the source emits the message $m$, we can define the {\it logical probability} that the source emits the symbol $x$ as
\begin{equation}
    p_S(x)=\sum_{\substack{m: x=f(m);\\m \in {\cal M}_s}} p_{{\cal M}_s}(m).
\end{equation}
The {\it semantic information} of symbol $x_i$ can be then defined as \cite{bao2011towards}:
\begin{equation}
    H_{\cal S}(x_i)=-\log_2 p_{{\cal S}}(x_i)
\end{equation}
The {\it semantic entropy} associated to the symbols emitted by the source is \cite{bao2011towards}
\begin{equation}
    H_{\cal S}(X)=- \sum_{x_i\in{\cal X}}p(x_i)\, \log_2 p_{{\cal S}}(x_i).
\end{equation}
In general, the two entropies $H_{\cal S}(M)$ and $H_{\cal S}(X)$  differ from eah other. Using basic information theory tools, we can always write
\begin{equation}
    H_{\cal S}(X)=H_{\cal S}(M)+H_{\cal S}(X/M)
    -H_{\cal S}(M/X),
\end{equation}
where $H_{\cal S}(X/M)$ denotes the entropy of $X$ conditioned to $M$ and $H_{\cal S}(M/X)$ denotes the entropy of $M$ conditioned to $X$: $H_{\cal S}(X/M)$ represents {\it semantic redundancy}, as it differs from zero only when there are multiple symbols associated to the same message, while $H_{\cal S}(M/X)$ denotes {\it semantic ambiguity}, as it differs from zero only when there are multiple messages (meanings) associated to the same symbol. Based on the above definitions, the goal of source (semantic) coding is {\it not} to preserve the sequence of symbols generated from the source, but  its semantics, i.e. the meaning associated to the emitted messages. To this end, it was shown in 
\cite{basu2014preserving} that there exists a semantic block encoder that only needs, on average, $I(M; X)$ bits to encode the messages emitted by the source, where 
\begin{equation}
    I(M; X)=H_{\cal S}(M)-H_{\cal S}(M/X)=H_{\cal S}(X)-H_{\cal S}(X/M)
\end{equation}
is the mutual information between source messages and source symbols. Some practical semantic source encoders were proposed in \cite{basu2014preserving}, exploiting shared knowledge between source and destination.\\

At the destination side, the received signal $y$ is syntactically decoded to produce a sequence  of symbols $x'$, which is then interpreted, based on the knowledge system ${\rm KB}_D$ available at destination, to provide a message $m'$. In principle, the two operations can be mixed, so that the goal of the semantic decoder is to recover, from $y$, a message $m'$ that is equivalent to $m$. 
We say that two messages are equivalent if they convey the same meaning. Equivalence does not necessarily imply that the structure of the message $m'$ associated to the concept is identical to the structure of $m$. What is necessary is only that, once interpreted according to the knowledge base system ${\rm KB}_D$ available at destination, the concept extracted from $m'$ be semantically equivalent to that represented by $m$.

In a semantic communication system, there might be errors at the syntactic level as well as errors at the semantic level: an error at syntactic level occurs if $x'$ differs from $x$; an error at semantic level means that $m'$ is not equivalent to $m$. Errors at syntactic level may occur because of the presence of random noise or interference introduced during the transfer through the channel, or because of unpredictable channel fluctuations. Errors at semantic level could be due to differences between the KB systems available at source and destination nodes, or because of misinterpretation.

Clearly, the semantic level relies upon the syntactic level: too many errors in the decoding of the received sequence $y$ may preclude the recovery of the source message $w$. However, and this is the interesting new aspect brought forward by the inclusion of the semantic level, {\it an error at the syntactic layer does not necessarily induce an error at the semantic layer}. The message interpreter can in fact recover the right content even if there are a few errors in decoding the received sequence of symbols. In other words, the semantic interpreter can correct a number of errors occurring at syntactic level exploiting the rules (logic) of the language subsuming the exchange of information.\\

Conversely, there could be errors at semantic level, even if there is no error at the syntactic level. This may happen, for example, when there are differences between the KB's available at source and destination, so that a message that has been correctly decoded at the syntactical level gets misinterpreted at the semantic level. 

Let us assume that  the channel is modeled through the conditional probability $p(y/x)$ of receiving a symbol $y$, once the symbol $x$ has been transmitted. If what matters is the recovery of the semantic message (meaning), rather than the corresponding symbol $x$, we can use a semantic decoder that chooses the message $m'$ that maximizes the posterior probability conditioned to the received symbol:
\begin{equation}
    m'={\rm arg}\hspace{-0.4cm}\max_{m: x=f(m)} p(m/y)={\rm arg}\hspace{-0.4cm}\max_{m: x=f(m)} \sum_x p(m, x, y).
\end{equation}
Using the Markov property that $p(y/m, x)=p(y/x)$, the decoding strategy can be rewritten as \cite{bao2011towards}
\begin{equation}
    m'={\rm arg}\hspace{-0.4cm}\max_{m: x=f(m)} \sum_x p(y/x) p(x/m) p(m).
\end{equation}
Since $p(m)$ and $p(y/x)$ are given, this formula says that the optimization of the overall system performance involves the search of the function $p(x/m)$ that minimizes the semantic error probability, given the constraints at the physical layer. The function $p(x/m)$ plays the role of a semantic encoder. Intuitively speaking, if there are not too many errors at syntactic level, we can expect a significant performance improvement resulting from semantic decoding because many received sentences can be corrected by exploiting the knowledge of the language (either human or artificial) used to communicate. As a trivial example, a simple spell checker that exploits the knowledge of the vocabulary and grammar used in the language adopted for text transmission can correct many misspelled words or sentences. In more abstract terms, with reference to Fig. \ref{message mapping}, what happens is that there could be many sentences, i.e. multiple points in the set ${\cal Y}$, which correspond to a single point (e.g., sequence $m$), in ${\cal M}_d$, associated to a semantically correct sentence.\\

\noindent{\bf Example of application to text transmission}\\

\begin{figure}[ht]
{\centering
\includegraphics[width=7.5cm]{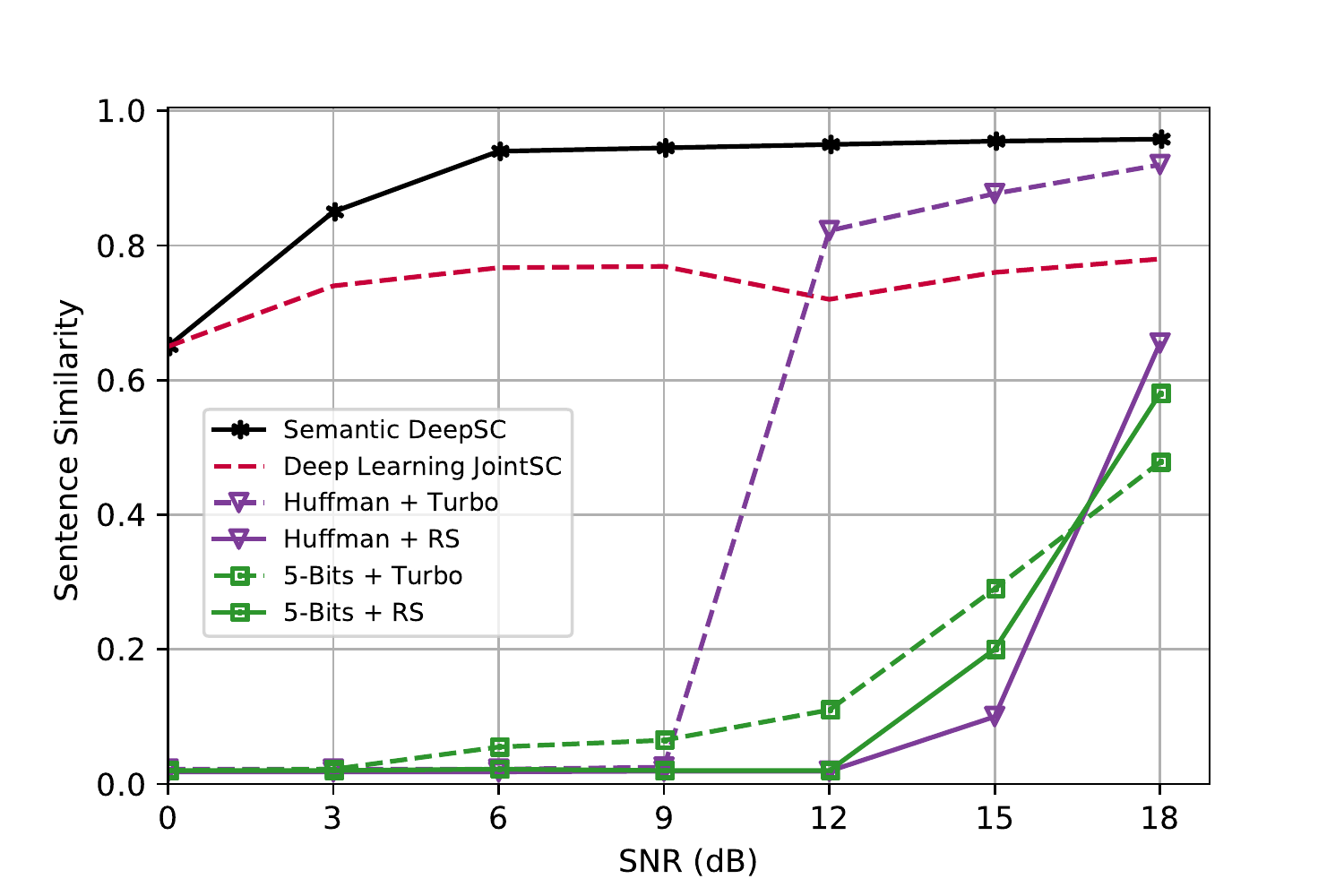} \includegraphics[width=7.5cm]{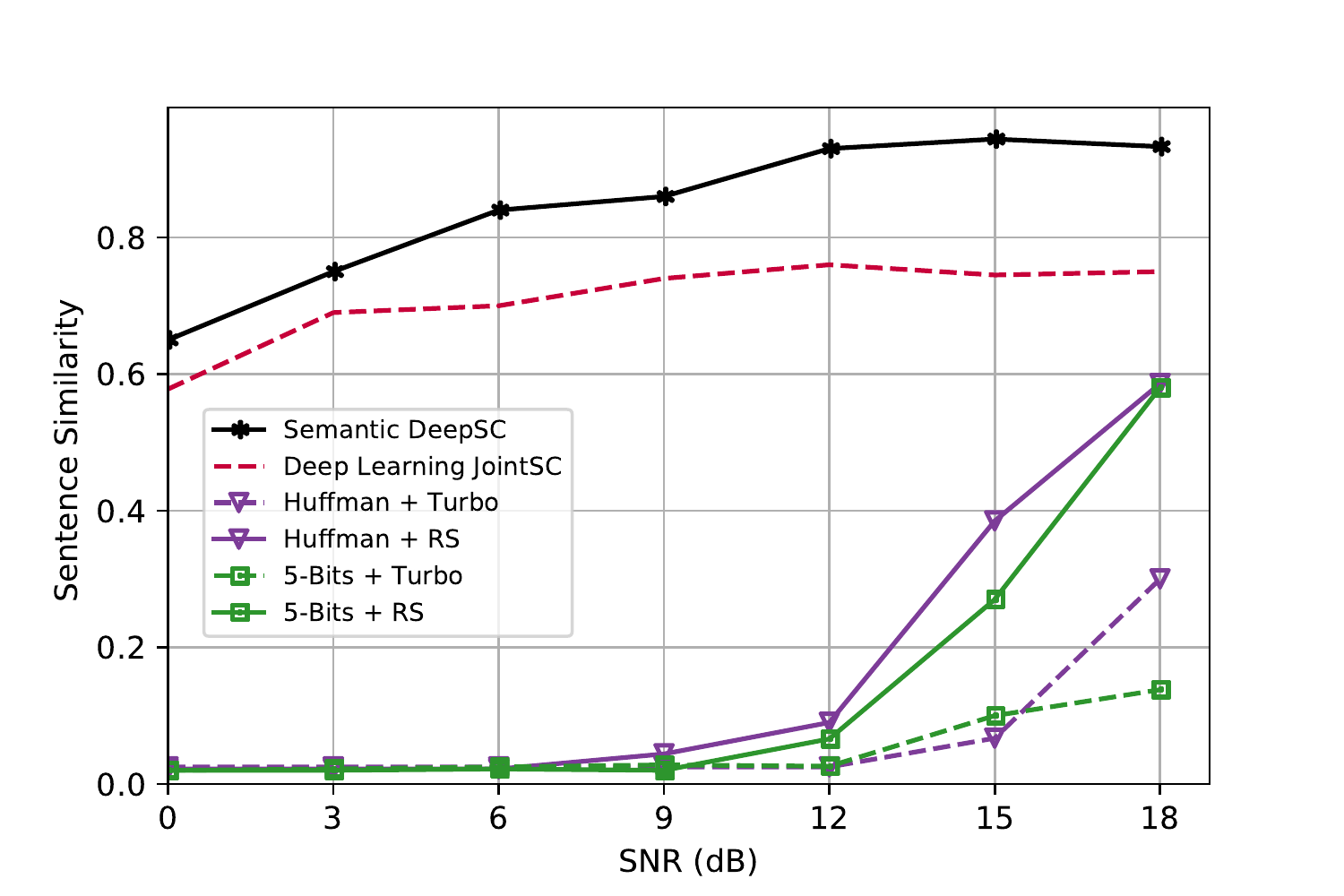}}\\
{\centering
{\small \hspace{2cm}a) AWGN channel; \hspace{4cm} b) Rayleigh fading channel}}
\caption{Semantic similarity vs. SNR (dB); courtesy of \cite{xie2020deep}.}
\label{DeepSC}
\end{figure}

In general, finding the optimal semantic encoder $p(x/m)$ is not an easy task. However, simpler yet effective alternatives are available. One example is given by the semantic communication scheme proposed in \cite{xie2020deep}, where the authors propose a deep learning based semantic method for text transmission, named DeepSC. The goal of DeepSC is to maximize the system capacity, while minimizing the semantic error by recovering the meaning of the sentences, rather than the bit-by-bit sequence of the transmitted signal. The DeepSC method is based on a deep neural network, which is trained with sentences of variable length to obtain a combined semantic/channel encoder. The objective of training is to minimize the semantic error, while reducing the number of symbols to be transmitted. Some numerical results, obtained in \cite{xie2020deep}, are reported in Fig. \ref{DeepSC} a) and b), referring, respectively, to an additive white Gaussian noise (AWGN) channel and to a Rayleigh fading channel. The performance is evaluated in terms of the similarity between the sentence emitted by the source and the sentence reconstructed by the semantic decoder at the receive side. The performance of DeepSC is compared with the following alternative methods:  Huffman coding with RS (30,42) in 64-QAM; 5-bit coding
with RS (42, 54) in 64-QAM; Huffman coding with Turbo coding in 64-QAM; 5-bit coding with Turbo coding in 128-QAM; an E2E DNN trained over the AWGN
channels and Rayleigh fading channels, proposed in \cite{farsad2018deep}. As we can see from Fig. \ref{DeepSC}, at low SNR values, where conventional schemes suffer from many errors at bit-level, the semantic decoder DeepSC significantly outperforms the alternative methods, exploiting the structure of a natural language. The method outperforms also the deep learning based method of \cite{farsad2018deep}, thanks to the exploitation of the semantic aspect. 
In this example, the reason why the semantic scheme outperforms all other methods is because it exploits the shared knowledge of the language used to communicate. This is an {\it a priori} knowledge, shared by source and destination, which does not need to be transmitted, but that imposes a structure on the interpreted words or sentences that helps to correct a lot of errors occurring at the physical layer.
\subsection{Feedback and cross-layer interaction in semantic communications}
Besides improving forward error correction, semantic decoding can also help significantly if employed in an Automatic Repeat reQuest (ARQ) scheme, using only error detection at the receiver. In fact, while a conventional receiver would require the retransmission of erroneous packets, at syntactic level, a semantic decoder would require a retransmission only if there is an error at the semantic level. The challenge, in this case, is to devise mechanisms to detect errors occurring at semantic levels. In principle, the interpreter will require a retransmission only when the recovered message The further possibility made possible by the new semantic framework is that the feedback from receiver to transmitter may now involve not only the sequence of transmitted symbols (syntactic level), but also the semantic aspect of the transmitted message, as depicted in Fig. \ref{Semantic-comm}.
A semantic feedback can be sent whenever the meaning of the message provided by the semantic interpreter is unclear. The message interpreter at destination can in fact send a feedback to the semantic message generator at the source side, to require the retransmission of the message $w$, or maybe a different version of $w$ that facilitates its interpretation at the receiver side. An additional feature of the semantic communication system is also the interaction between semantic and syntactic  levels. The semantic interpreter at the destination can in fact send a feedback to the syntactic encoder as well, as shown in Fig. \ref{Semantic-comm}. For example, the semantic decoder can tell the source encoder to reduce the data rate because the message that is being received can be easily decoded and predicted (up to a certain time interval), at the semantic level, so that it is not necessary to transmit all the fine details that are currently sent. In this way, the whole system might achieve the same accuracy in the recovery of the transmitted information, but saving important physical resources such as energy or bandwidth.\\

In summary, the interaction between different communication levels paves the way to a new way to design communication systems. Nowadays, communications are designed to ensure that there are no errors at syntactic level, which means that the sequence of symbols used to encode the message emitted by the source should be correctly received at destination, irrespective of what is being transmitted, i.e. of the information content encoded in the transmitted message. In a semantic communication system, what matters is that the receiver be able to recover the content of the information sent by the source. There could be errors at syntactic level that could be easily corrected at the semantic level, without requiring the retransmission of the corresponding packets. Going even further, there may be parts of the message that may not be able to reach the destination, perhaps due to blocking effects at the physical level, such as in millimeter wave communications, but the interpreter may still be able to reconstruct the semantic message, based on a well tuned prediction model. Where do these advantages of semantic communications come from ? In a nutshell, the fundamental gain comes from the fact that, typically, source and destination share a lot of common knowledge. This shared knowledge is what makes possible to correct many errors or to avoid sending details that can be easily recovered from the context. The price paid for these advantages is additional receiver complexity.

\subsection{Examples of applications of semantic communications}
The example shown before on text transmission is helpful to introduce the concept of semantic communication, but it involves a low data rate application for which there is no urgent need to improve performance. In this section, we highlight three representative examples of applications of increasing challenge and difficulty, which could significantly benefit from the introduction of semantic aspects. 
\subsubsection{Speech signals}
The transmission of speech signals can benefit from the incorporation of NLP tools and the incorporation of a speech recognition step that translates speech into text. The automatic correction of words, or sentences, can in fact lower the pressure on forward error correction codes because more errors at bit-level can be compensated by the word/sentence correction mechanisms. Alternatively, pieces of speech that get lost because of a deep channel fade can be reconstructed from the context, without requiring their retransmission.

\subsubsection{Video streaming} Video streaming is already consuming most network resources, so this is a use-case where an improvement is more expected. The question is how to extend the semantic framework to deal with video signals as well. In our view, this is possible by incorporating an interpretation of what is going on in the video. 
Suppose, for example, that the video is capturing the scene of a walking person and that, at some points, a number of frames are lost because of a deep channel fade. An interpreter, at the receive side, could segment the video and distinguish the walking person from the background. Building on previous frames, the interpreter can predict the next frames, using a properly trained prediction model. Recent examples of video coding incorporating frame prediction based on deep neural networks are presented in \cite{choi2019deep}. If no major unexpected change occurs during the channel fade, the overall flow of events captured by the video could be reconstructed, with apparently no harm at semantic level. In such a case, the message interpreter would be able to reproduce a video that is not necessarily equal to the transmitted video, but it is semantically equivalent. Clearly, this approach can produce a significant saving in terms of transmit power and/or bandwidth. 

\subsubsection{Holographic communications} A challenging new use cases foreseen for 6G is holographic communications, where multiple views of a scene are transmitted to enable the creation of a hologram at the receiver side. In this case, significant advances can be expected from the incorporation of semantic features, whenever source and destination share some common knowledge background. As an example, suppose there is a speaker delivering a talk at a remote side, where she is represented as a hologram. The receiver could have a digital model of the speaker. Exploiting the model, the receiver could reconstruct some of the multiple views necessary to reconstruct the hologram in real time, with evident benefits in terms of data rate strictly necessary to provide the desired quality of experience.

Further semantic communication based applications are possible to serve future challenging use cases such as brain-to-computer interaction, multi-sense and multi-sensory XR, affective computing, intertwining (natural and/or artificial) intelligence, etc. (see Table \ref{tab:6G-Use Cases}).

The price paid for the advantages offered by the inclusion of semantic aspects is the additional computational complexity at the receive side and, in turn, a further delay, which could represent,  in some applications, a serious bottleneck. To reduce the additional delay, in our vision, future communication systems could take valuable suggestions from the observation of how human brains operate. Among the many theories of human mind, there is a beautiful theory denoted as {\it predictive mind}, supported by experimental evidence  \cite{hohwy2013predictive}, \cite{clark2015surfing}. According to this theory, the brain is a bundle of cells that support perception and action by constantly attempting to match incoming sensory inputs with top-down expectations or predictions, learning from prediction errors. A striking example is vision \cite{clark2013whatever}. The brain is continuously creating an image of the outside world, based on what already knows and on what it observes, using a hierarchical generative model that aims to minimize prediction error within a bidirectional cascade of cortical processing. According to this theory, it is the brain that selects a small subset of the multitude of signals coming from the retina, as a function of what it is expecting. In this way, most signals produced in the retina do not travel through the optical nerve. There is a significant flow of information from the retina to the brain only when the observation deviates significantly from the prediction. This represents indeed a very efficient way of working, as it saves a lot of energy, and it could be translated into next generation artificial visual systems. 


\section{Goal-oriented communication}
\label{Goal-oriented communication}
In this section, we explore some of the possibilities offered by the inclusion of the effectiveness level in the protocol stack, as shown in Fig. \ref{architecture}. We focus on the case in which effectiveness is achieved by specifying a clear goal of the communication. In such a case, the idea is to transmit not all the information, but only the information that is strictly relevant to the effective fulfillment of the goal, optimizing the system performance while satisfying the constraints dictated by the application. We denote this framework as  {\it goal-oriented} communication. 
Earlier works on goal-oriented communications are \cite{juba2008universal}, \cite{juba2011universal} and its extension \cite{goldreich2012theory}. In those works, the authors addressed the problem of potential “misunderstanding” among parties involved in a communication, where the
misunderstanding arises from lack of initial agreement on what protocol and/or language is being used in communication. In this section, we propose a different view, starting from the basic assumption that the communication occurs to fulfil a  goal.
As a consequence, the performance of the system is specified by the degree of fulfillment of the given goal or, more precisely,  on the {\it effectiveness} achievable in the fulfillment of the goal given the amount of resources used to do it.
Let us consider, for the sake of clarity, the case in which the goal of the communication is to estimate the parameters of a function modeling the data collected by a set of sensors. The problem is to  evaluate the minimum number of bits to be transmitted strictly necessary to enable the fusion center to estimate the set of parameters with the required accuracy. This problem has been addressed in several works.
In particular, the problem is relevant in the emerging field  of edge learning, where learning tools are brought closer and closer to the end user to meet stringent delay constraints \cite{park2018wireless}. In the edge learning framework, having in mind the goal of learning and the resources dedicated to that goal, several trade-offs are possible, like the trade-off between power consumption and delay, between accuracy and delay, etc. The authors of \cite{Skatchkovsky2019} consider an edge machine learning system, where an edge processor runs an algorithm based on stochastic gradient descent (SGD), to reach a trade-off between latency and accuracy, by optimizing the packet payload size, given the overhead of each data packet transmission and the ratio between the computation and the communication rates. In \cite{Mohammad2019}, the authors proposed an algorithm to maximize the learning accuracy under latency constraints. In  \cite{Amir2019}, it was proposed a distributed machine learning algorithm at the edge, where wireless devices collaboratively minimize an empirical loss function with the help of a remote server.

\begin{figure}[ht]
\centering
\includegraphics[width=\columnwidth]{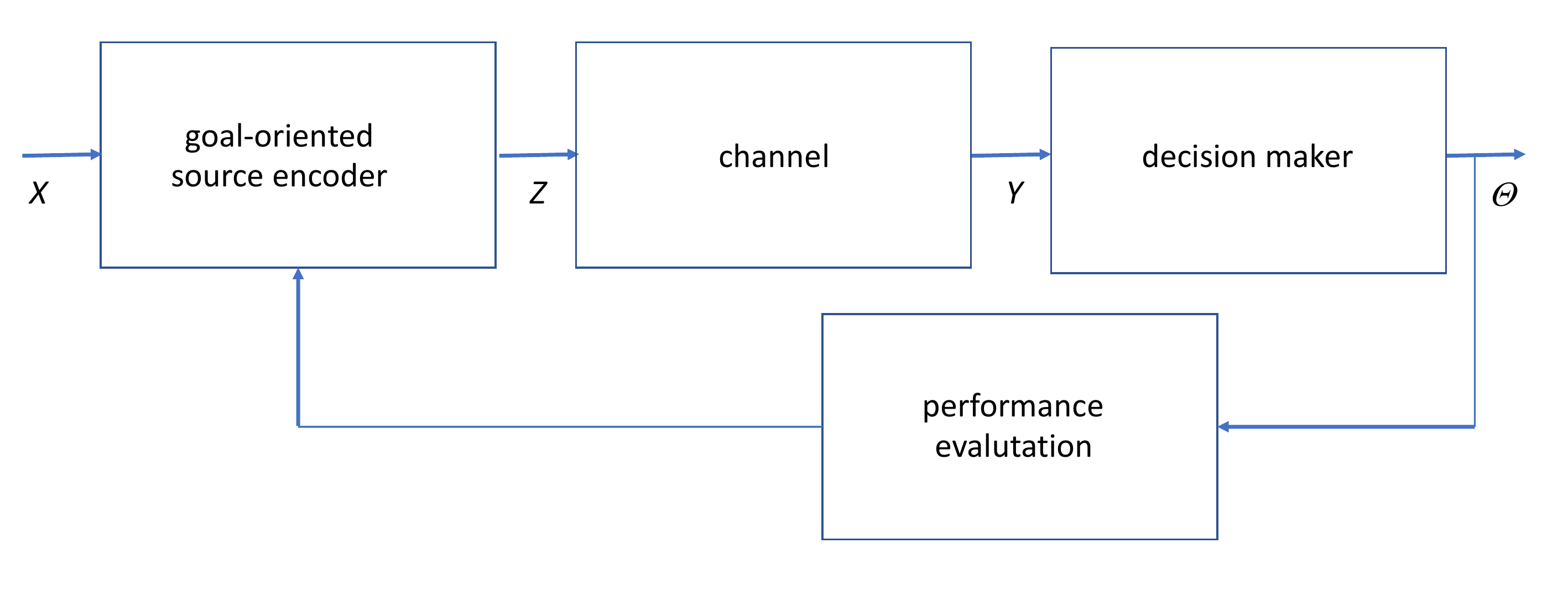}
\caption{Goal-oriented communication.}
\label{GO-scheme}
\end{figure}

In this section, we propose a general framework to address the problem.
The system can be sketched as in Fig. \ref{GO-scheme}, where the novelty is represented by the feedback from the decision maker to the source encoder.
To present the general idea, let us suppose that the goal of the communication system is either to perform classification or to learn a set of parameters $\boldsymbol{\theta}$ from a set of observations $\mathbf{x}_i, i=1, \ldots, N$. Let us denote with $\mathbf{X}:=\{\mathbf{x}_i\}_{i=1}^{N}$ the set of measurements. The goal is to send the data to a fusion center that takes a decision. Let us denote  by $\mathbf{Z}$ the data sent to the fusion center. The question is how to encode the data, i.e. how to map $\mathbf{X}$ into $\mathbf{Z}$, to satisfy the service requirements, e.g. service delay and decision accuracy,  while possibly minimizing energy consumption. Typically, the data are source encoded to remove redundancy, still allowing perfect reconstruction (lossless compression) or to reach a satisfactory compromise between distortion and encoding rate (lossy compression) \cite{cover1999elements}. We propose a different strategy. The idea is that, if communication occurs to fulfil a goal, for example parameter estimation or classification, the source encoder should be designed in order to  achieve the desired accuracy on the parameter estimation or classification, while minimizing the use of physical resources or to minimize the time needed to take a decision. The source encoder should then be tuned according to the performance on the learner operating at the fusion center, rather than minimizing distortion in the reconstruction of the observed data. Let us suppose that the parameter vector to be estimated is itself a vector random variable, denoted as $\mathbf{\Theta}$. We can cast the problem in a formal way, as the search of the mapping $\mathbf{Z}=\mathbf{f}(\mathbf{X})$ such that $\mathbf{Z}$ is maximally compressed, but there is no information loss when passing from $\mathbf{X}$ to $\mathbf{Z}$, as far as the recovery of $\mathbf{\Theta}$ is concerned, i.e.
\begin{equation}
    I(\mathbf{X}; \mathbf{\Theta})=I(\mathbf{Z}; \mathbf{\Theta}),
\end{equation}
where $I(\mathbf{X}; \mathbf{Y})$ denotes the mutual information from $\mathbf{X}$ to $\mathbf{Y}$ \cite{cover1999elements}. The solution to this problem is given by the {\it minimal sufficient statistics} of $\mathbf{X}$ \cite{cover1999elements}.
From basic statistical signal processing, we know that $\mathbf{f}(\mathbf{x})$ is a sufficient statistics for $\boldsymbol{\theta}$, given $\mathbf{x}$, if the joint pdf $p(\mathbf{x}; \boldsymbol{\theta})$ can be factorized as \cite{cover1999elements}:
\begin{equation}
    p(\mathbf{x}; \boldsymbol{\theta})=g_{\boldsymbol{\theta}}[\mathbf{f}(\mathbf{x})]\, h(\mathbf{x}).
\end{equation}
In general, there might exist more than one sufficient statistic. 
What is important in our setting is to identify the {\it minimal} sufficient statistics. A sufficient statistic $\mathbf{f}(\mathbf{x})$ is minimal, relative to $p(\mathbf{x}; \boldsymbol{\theta})$, if it is a function of every other sufficient statistic \cite{cover1999elements}. In words, a minimal sufficient statistic maximally compresses the information about the vector parameter $\boldsymbol{\theta}$ in the observed samples. 

The above formulation means that, if the goal of communication is estimating the parameter vector $\boldsymbol{\theta}$ from the data-set $\mathbf{X}$, there is no loss of information in sending $\mathbf{f}(\mathbf{X})$ instead of $\mathbf{X}$. What is the advantage? The advantage is that the entropy of $\mathbf{f}(\mathbf{X})$ could be much smaller than the entropy of $\mathbf{X}$. This means that the number of bits necessary to encode $\mathbf{f}(\mathbf{X})$ may be much smaller than the number of bits necessary to encode $\mathbf{X}$. As a consequence, the number of bits to be transmitted can be significantly decreased, with no losses in terms of inference. 

The formulation given above exploits the sufficient statistic of the data. In some cases, it may be difficult, or even impossible, to find a minimal sufficient statistics. To overcome these difficulties, \textcolor{black}{we propose a very general approach, based on the information bottleneck principle  \cite{tishby2000information}.} The approach can be formulated as follows: Search for the encoding function $\mathbf{Z}=\mathbf{f}(\mathbf{X})$ that achieves the best trade-off between the loss of information about the estimation of $\mathbf{\Theta}$, resulting from compressing the input $\mathbf{X}$, and the number of bits necessary to encode $\mathbf{f}(\mathbf{X})$. The problem can be formulated as
\begin{equation}
    \min_{\mathbf{f}(\mathbf{X})} I(\mathbf{X}; \mathbf{Z})-\beta I(\mathbf{Z}; \mathbf{\Theta}),
    \label{bottleneck}
\end{equation}
where $\beta$ is a positive real number used to assign different weights to the two terms in \eqref{bottleneck}: A small value of $\beta$ privileges maximum compression of the source data, at the expense of learning accuracy, whereas large values of $\beta$ privilege the learning accuracy, at the expense of source compression. Even though the above problem looks prohibitive, if the involved probability density functions are given, its solution can be achieved through an iterative algorithm, known as the {\it information bottleneck} method \cite{tishby2000information}. More recently, the method has been used to explain the learning and generalization capabilities of deep neural networks, thus showing its deep impact in learning problems \cite{shamir2010learning}. Finding the optimal encoder according to the information bottleneck may be difficult in the general case, where the statistics of the data are unknown. However, learning structures like auto-encoders can provide a viable good approximation. In our formulation, the parameter $\beta$ present in \eqref{bottleneck} can be adapted online, depending on the channel state and on the level of accuracy achieved by the learner, in order to find, dynamically, the best compromise between energy consumption, E2E delay and learning accuracy.\\

\noindent
{\bf Example of application:   Edge online learning}\\
We show now a simple example of application of a goal-oriented system that takes into account the goal of the specific goal of communications. In this example, we consider the case of an automated image classification problem.
\begin{figure}[ht]
\centering
\includegraphics[width=10cm]{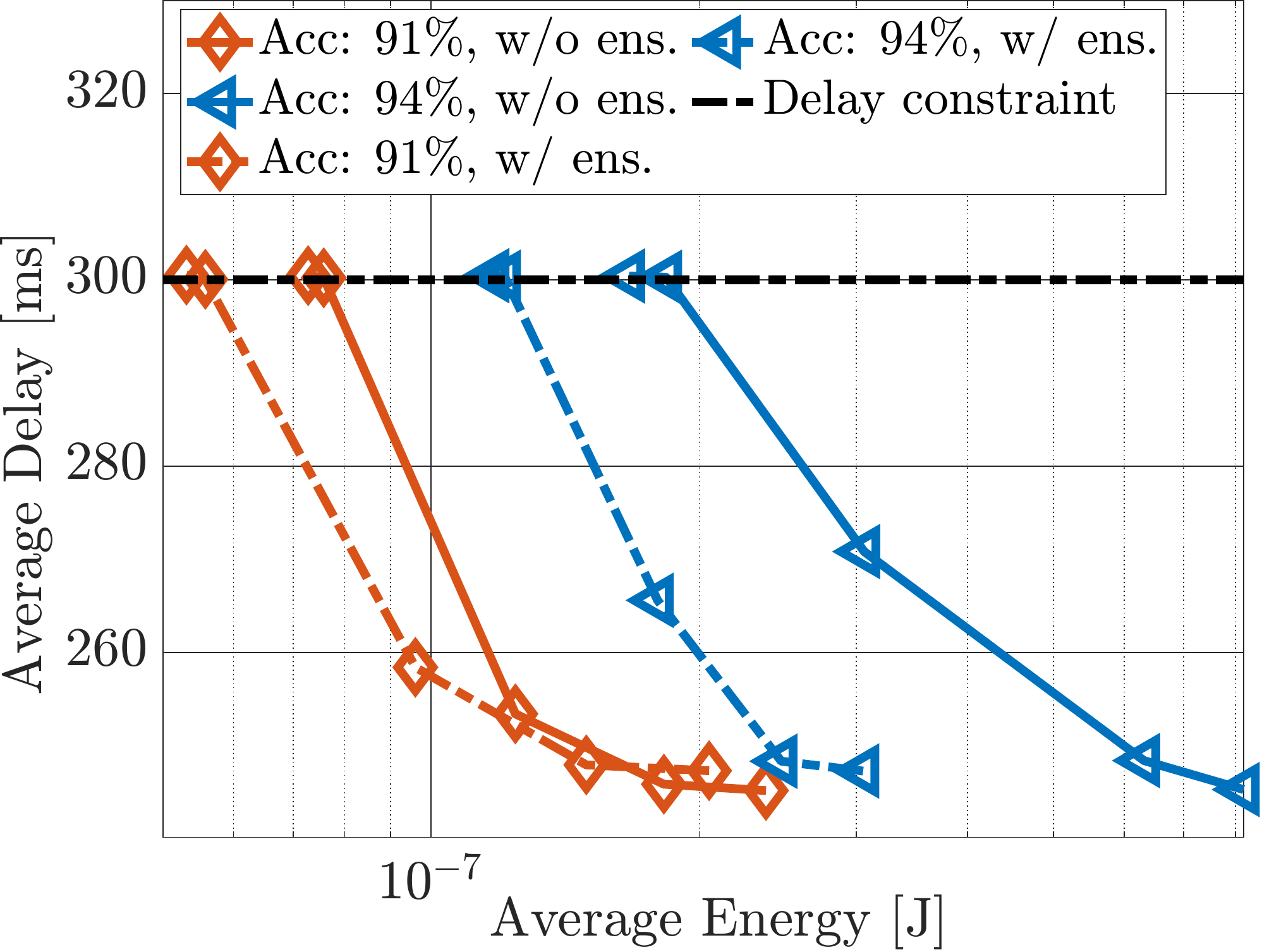}
\caption{Optimal energy-delay trade-off necessary to achieve a desired classification accuracy.}
\label{tradeoff-ensemble}
\end{figure}
The set-up consists of a sensor that collects data and sends \textcolor{black}{it} to an edge server that trains a classifier. The proposed method minimizes a weighted sum of energy consumption and classification accuracy, under constraints on the average E2E delay. In particular, the method dynamically adjusts the number of quantization bits, the transmit power, and the CPU scheduling at the edge server. The method has been tested on the MNIST dataset \cite{726791} and on the Hydraulic System Monitoring (HSM) dataset \cite{7151267}. To increase the reliability of the classifier, an {\it ensemble} method is adopted, combining multiple classifiers (learners) that are trained in parallel on the same set of data. The decision is then taken from the learner that yields the most reliable output.  In this specific example, there are two  learners running in parallel on the same data: a Support Vector Machine (SVMs) \cite{scholkopf2000new} and a standard MultiLayer Perceptron (MLP). The data are sent through a wireless channel affected by fading. The method adjusts dynamically the number of bits used to encode the images, in order to satisfy a constraint on the E2E delay and strike the best tradeoff between classification accuracy and  energy consumption of the sensor. The adaptation of the source encoders exploits the feedback from the decision maker at the receiver side, evaluated in terms of classification accuracy, to the source encoder at the transmitter side. Since the correct classification cannot be really evaluated in practice, we use as a heuristic the entropy computed on the outputs of the SVM or the MLP, properly normalized to represent the a posteriori probabilities of a class, conditioned to the observation. Low values of entropy correspond to the case in which the classifier shows a strong preference for one class; large values of entropy represent a total confusion of the classifier. The running value of the entropy is also exploited to switch, online, from one classifier to the other.   The performance is shown in  Fig. \ref{tradeoff-ensemble}, representing the best trade-off between energy spent by the transmitting nodes and the delay necessary to achieve the desired learning accuracy. From the figure, we can also observe the gain achieved combining multiple learners trained in parallel. The two sets of curves refer to two different cases where the objective function to be minimized represents a different combination of energy and E2E delay. For each case, we report the probability of correct classification. In each case, we consider the ensemble learning, combining two learners (w/ensemble) and a single learner (w/o ensemble). Comparing the figures we can see how, properly setting the parameters of the system, we can get a substantial improvement in the overall trade-off between energy, E2E delay and learning accuracy. In particular, we may notice how, passing from a high level of accuracy ($94 \%$ - blue curves) to a slightly lower value ($91 \%$ - red curves), we can achieve a substantial energy saving, for the same value of E2E delay.\\

In the above simple example, the source encoder is adjusted acting on the number of quantization bits. Clearly, more sophisticated techniques are possible. A powerful general purpose approach is offered by deep neural networks (DNN), used to make inference on the observed data. In such a case, we can measure the mutual information between the input and each internal layer of the network. Experimental results reveal that the first layers of a DNN operate some kind of compression over the course of training \cite{goldfeld2018estimating}. This reduction
is driven by progressive geometric clustering. Building on such a behavior, we can think of splitting the DNN in a certain number of layers, putting the initial layers, involved in some kind of compression or feature extraction, at the source and the remaining layers at the destination, so as to reduce the number of bits to be transmitted from source to destination, still maintaining roughly the same overall performance. 

A use case that is gaining increasing attention in edge computing is the extraction of video analytics, possibly in real time. In such a case, the definition of a goal can help to reduce the data rate from peripheral video camera and the edge server where the video analytics are extracted. One possibility is, for example, to perform a preliminary filtering to remove all frames that are not relevant for the ensuing video analysis. As an example, FFS-VA is a pipelined system for multistage video analytics, based on three stages \cite{zhang2018ffs}: an initial stage used to remove the frames only containing a background; a stream-specialized network model used to identify target-object frames; a model to remove the frames whose target objects are fewer than a threshold. 

The last example suggests that, if we merge semantic and goal-oriented communications, we may find methods to send only the semantic information that is strictly  relevant to the achievement of the goal. In this way, we have a further way to reduce the amount of data to be sent, without affecting the accuracy/reliability in the goal accomplishment.

\section{Online Learning-Based Communication and Control}
\label{Learning-based design in a C4 framework}
The grand vision beyond 6G is that, whereas all previous network generations have been designed by humans, the design of next generation networks will see a significant contribution from machines, driven by a pervasive introduction of artificial intelligence at the edge of the network, as close as possible to the end-users \cite{zhou2019edge}, \cite{peltonen20206g}. Bringing intelligence at the edge of the network meets a twofold request \cite{park2018wireless}: {\bf Machine Learning for Communication, Computation, Caching, and Control}, aimed to optimize the utilization of network resources, distributed contents and implement the appropriate control actions, by learning network-related parameters, content popularity and predicting future events; and {\bf Communication for Machine Learning}, aimed to exploit communication to enable the distributed implementation of machine learning algorithms enabling context-aware delay critical services. To support this view, 6G will have to be an {\it AI-native} network, meaning that the network will be designed to facilitate the introduction of learning tools that will reshape the network according to requirements and constraints \cite{letaief2019roadmap}. 

In this paper, we support this vision while stressing, at the same time, that \textcolor{black}{a true leap forward can be achieved by merging machine learning, which builds on data-driven inductive strategies, with semantics, which is based on model-based deductive strategies.} The interplay between learning and semantics will produce a twofold advantage: i) semantic communications, with its widespread use of knowledge representation systems, will facilitate the development of machine learning algorithms that exploit semantic features to improve their learning capabilities, facilitate disambiguation exploiting context information, and increase robustness against adversarial attacks; ii) machine learning will help semantic communications algorithms to better understand which is the relevant information, thus further improving effectiveness/efficiency. In the following sections, we will elaborate on the above ideas.

\subsection{Holistic Orchestration of $C^4$ Resources}

Introducing intelligence to enable new services, like for example intelligent manufacturing, autonomous driving, or virtual reality, to cite a few, requires taking smart decisions within tight delay constraints \textcolor{black}{and to respect the jitter bound to enforce a deterministic chain decision process}. To meet this demand, edge computing represents an emerging paradigm that pushes computing tasks and services from the (possibly distant) cloud to the edge of the network. An efficient design of edge computing should enable the end users, either humans or machines, to access computational and storage resources with very low service delays. 

The service delay, i.e. the time elapsed between the instantiation of a request and its fulfillment, typically involves a communication delay between the involved parties, a computation delay, and possibly the time needed to access storage units containing relevant data. Furthermore, if the goal of communication is the control and actuation of delay-critical procedures, it is also necessary to include a further delay associated to control and actuation. As a consequence, imposing a service delay induces a coupling between communication, computation, caching, and control ($C^4$) resources. This coupling motivates a joint $C^4$ design, to achieve an effective resource orchestration \cite{barbarossa2018edge}. In this $C^4$ context, \textcolor{black}{service placement and request routing plays a key role in distributing computational resources across the networks}. A further key service is computation offloading, which plays a key role in extending the capabilities of peripheral devices to take smart decisions, by offloading their computational requests to nearby mobile edge hosts (MEH).  \textcolor{black}{In \cite{Ndikumana2020ieeeTMCC4}, \cite{Ndikumana2019C4PhD} the authors propose a joint $C^4$ model for collaborative MEC, while in \cite{Wang2019INFOCOM-C4} the authors propose an edge controller that \textcolor{black}{implements an optimal control strategy for connected cruise control in advanced driver assistance systems, relying on the joint optimization of  communication, computation, and caching resources}. In the above mentioned examples, the delay-criticality  and the consequent associated energy consumption of received and processed data represents a still unresolved bottleneck.}

To satisfy economy of scale issues, the network design must keep into account that edge resources, like computing and storage capabilities, will be necessarily limited, so that it is imperative to optimize their usage. For example, in computation offloading the service delay includes a communication delay and a computation delay. Hence, it makes sense to optimize the use of computation and communication resources {\it jointly}, as suggested for example in \cite{sardellitti2015joint}, in a static multi-user setup, where multiple small cells are served by a single edge computing host. A dynamic joint optimization algorithm that schedules the communication and computational resources optimally was suggested in \cite{mao2017stochastic}, \cite{merluzzi2020dynamic}. A further extension to the case where, in each slot, the optimizer may not know exactly the state of the system and then of the objective function to be optimized, one may resort to online convex optimization (OCO) algorithms, as suggested in \cite{chen2019learning}. 
Besides communication and computation, a further coupling in the $C^4$ framework is between communication and caching. In many applications, it is necessary to cache the desired contents {\it on demand} and, possibly, in a proactive way, to meet delay constraints. Proactive caching will be considered in Section \ref{Intelligent content delivery at the edge}. The additional novelty is that, in the $C^4$ context, caching will involve not only popular content, but also software needed to run user applications remotely, but as close as possible to the end-user.

\subsection{Machine learning for wireless networks}
Proactive mechanisms require the availability of prediction mechanisms and then, more generally, sophisticated machine learning strategies, possibly running at the edge of the network. 
In the last decade, in the field of supervised learning, deep neural networks (DNN) have been shown to provide performance even better than human capabilities, in some general purpose applications, like sound recognition or image classification \cite{lecun2015deep}. In particular, convolutional neural networks (CNN) are a key actor, for their intrinsic sparsification of the number of edges from one layer to the next, achieved exploiting the structure of convolutional operators, well suited for images and sound, i.e. signals residing over regular grids. More recently, the objective has been the extension of DNN architectures to work on data that do not reside over a regular grid, but rather on graphs, whose topology captures some of the intrinsic (pairwise) relations among the observed signals. Mixing graph-based input representation with learning is a problem addressed in \cite{bronstein2017geometric}. 

When applying supervised learning methods to communication networks, 
there is the great opportunity to train the \textcolor{black}{learner using a simulator that generates input data, channels, and corresponding outputs, using consolidated statistical  models. In this way, the learner can be trained using a huge number of labeled examples.} 

In supervised learning, there is typically a clear separation between the learning and testing phases. However, in a wireless context where the channel is going to vary over time, it is more interesting to look for {\it online} learning mechanisms, where the learning and testing phases  are intertwined and evolve over time. Among online algorithms, we may distinguish between reinforcement learning methods, where an agent learns by acting and observing the results of its action, without assuming an a priori model of the observation, and stochastic optimization methods, where a dynamic procedure is enforced to adapts its online actions, derived from an online optimization that, step-by-step, exploits whatever knowledge is available at the moment about the involved variables. 

In general, when using supervised learning, the price for achieving a high performance level is the need of a huge number of labelled data. This means that, typically, in supervised learning,  humans are still playing a fundamental role in providing the labelled examples. 
Conversely, in unsupervised learning, there are no labelled examples to use. In such a case, the goal of the learner is to find patterns in the data, like for examples clusters, and then classify the observation according to the features of the detected patterns. Graph-based representation methods are a fundamental tool for clustering, as evidenced in spectral clustering methods. However, graph-based approaches only capture {\it pairwise} relations among entities, like e.g., time series. However, in many applications, pairwise relations are not able to extract all the information. A further advancement towards the incorporation of {\it  multiway} relations has been carried out in \cite{barbarossa2020topological}, introducing topological signal processing (TSP). \textcolor{black}{TSP can be especially valuable when the observations are associated to the edges of a graph, like for example traffic data. In \cite{barbarossa2020topological}, it was shown how the use of higher order topological models (simplicial complexes) yield advantages over graphical models in predicting traffic maps from sparse measurements .}   Merging TSP with DNN has the potential to unravel important information from complex data sets. 

Examples of application of machine learning tools to the physical layer of a communication system have been already studied, under unknown channels, using an autoencoder (AE) \cite{dorner2017deep}, \cite{balevi2019one}, a recurrent neural network (RNN) \cite{farsad2018neural}, and a generative adversarial network (GAN) \cite{ye2018channel}. Extensions to higher layers, including network slicing and orchestration are presented in \cite{dandachi2019artificial}, where the authors propose an AI framework for cross-slice admission and congestion control considers communication, computing, and storage resources  simultaneously to maximize resources utilization and operator revenue.

\subsection{Federated learning}
The pervasive introduction of learning tools at the edge of the network poses a number of challenges. In a framework where multiple devices produce a wealth of data to be used to extract analytics, it is clear that some sort of collaborative learning can boost the performance of learning algorithms. Collaborative learning typically requires the exchange of data, but this approach raises critical issues because of privacy concerns to share data among users. A viable approach that has recently received a significant interest is {\it federated learning} \cite{yang2019federated}, \cite{li2020federated}, where learning the model parameters is performed over a central unit, either a data center or an edge host, while the data are kept in the peripheral nodes. In centralized federated learning, the devices do not send their data to any remote server. They only share local estimates of the parameters to be learned. Under quite broad assumptions, each device can boost its performance also without exchanging data, thus preserving privacy. A possible formulation of the objective function to be minimized in federated learning is the following
\begin{equation}
    \min_{\mathbf{w}} \sum_{i=1}^{N} p_i\,f_i(\mathbf{x_i}; \mathbf{w})
\end{equation}
where $f_i(\mathbf{x_i}; \mathbf{w})$ is the empirical loss function of device $i$, $\mathbf{w}$ is the (global) parameter vector to be learnt (e.g., a regressor or the wights of a DNN or the result of a classification),  $\mathbf{x}_i$ are the data collected by device $i$, and $p_i$, with $p_i\ge 0$ and $\sum_{i=1}^{N} p_i=1$, is a coefficient that weights the importance of the data collected by user $i$ in the estimation of $\mathbf{w}$. In the simplest setting, the weights $p_i$ can be chosen as $p_i=n_i/(\sum_{i=1}^{N}n_i)$, where $n_i$ is the number of examples observed by the $i$th machine. In federated learning, an iterative procedure is implemented where, at each iteration $n$, instead of sending the local data $\mathbf{x_i}$, each device sends its local estimate $\hat{\mathbf{w}}_i[n]$ (or a gradient of its local empirical loss with respect to the parameter vector) to a fusion center, which sends back an updating term that takes into account the information received by all cooperating nodes. Under convexity conditions about the optimization problem, this strategy converges to the global optimum \cite{li2020federated}. 

The above setting is amenable for its simplicity, but it also faces a number of challenges, namely heterogeneity in communication channels, local devices' behaviors, and models. More specifically, in a practical setting, the communication channels between the local devices and the fusion center may vary significantly across devices, in terms of data rates, latency and blocking probability. 
This heterogeneity alters the updating rule at the fusion center and then it can impact the final accuracy and the convergence time. Similarly, some devices can be faulty or provide data with high delays that again impact the convergence time. Finally, there is a model heterogeneity implying that there is no single globally optimal estimate $\mathbf{w}$ fitting all local needs, but there are rather different devices, or groups of devices, for which there is a better estimate $\mathbf{w}_k$, which does not necessarily coincide with the best estimate of another group of devices. In such a case, a valid improvement is represented by {\it multi-task} federated learning \cite{smith2017federated}.
\subsection{Intelligent content delivery at the edge}
\label{Intelligent content delivery at the edge}
\textcolor{black}{Dynamic caching is expected to play a key role to reduce the time elapsed between the instantiation of a request and the content delivery.} In edge-caching-based networks, a large amount of popular content can be pre-fetched and stored by the edge facilities, such as access points or mobile edge hosts, making a substantial portion of the data visible, ubiquitous and very close to the UEs. In particular, proactive caching policies, populating local storage disks based on estimated demand, is a key enabler. The optimization variables in dynamic caching are \cite{paschos2019cache}: cache deployment, deciding where to deploy the caches; content caching, deciding which files are put in  each cache; and content routing, deciding which paths will be employed to carry the right content to the right place. 
Clearly, the goodness of a proactive caching policy relies upon the accuracy of the prediction algorithms, which need to incorporate the variability of content requests over space ad time. So far, caching policies have been fundamentally addressed to move contents throughout the network. In the semantic communication framework highlighted in this work, what is necessary to move is not only content but also knowledge base systems and virtual machines, able to run applications on demand, as close to the end user as possible. Migrating virtual machines is a topic that has received significant attention, but a big leap forward  is still needed to reduce the migration times. This involves, for example, the use of {\it light} virtual machines, like containers, to reduce the amount of data to be migrated. In general, proactive caching of virtual machines poses an interesting and challenging problem in terms of computation and caching, especially for delay-sensitive applications. More generally, the distributed implementation of machine learning algorithms accessing distributed contents poses a number of challenges, like, e.g., namely communication bandwidth, straggler’s (i.e., slow or failing nodes) delay, privacy and security bottlenecks. A new concept that can alleviate some of the above bottlenecks in large-scale distributed computing is {\it coded computing}, which utilizes coding theory to effectively inject and leverage data/computation redundancy \cite{li2020coded}. More specifically, a method called {\it Coded Distributed Computing (CDC)} has been recently proposed in \cite{li2020coded}, which injects redundant computations across the network in a structured manner. 
\subsection{Semantic machine learning}
Machine learning is a data-driven approach that learns  and uncovers patterns from examples. Typically, learning is the result of an inductive  (bottom-up) process. Conversely, humans learn (abstract) models from experience and from the culture accumulated through time from human kind. They use these abstract models to interpret what they perceive, to plan actions, build further models, check their validity, and so on. Being purely inductive, machine learning tools are apparently unbiased. However, the patterns learned by machines sometimes are only brittle surface-level observational visual characteristics of the observed data. In object recognition, for example, typically the learner segments the image to better interpret it. However, sometimes it is precisely this act of segmenting that,  taking things out of context, may induce ambiguities. This is one of the reasons why current deep learning systems may sometimes be so brittle and easy to fool despite their uncanny power \cite{fawzi2015fundamental}: They search for correlations in data, rather than meaning, but meaning is much more than correlation.

Conversely, humans put their observations into context, thus connecting their observations to previously accumulated knowledge and properly reasoning about what they are sensing, always struggling to make sense, i.e. extract meaning, from their perception.


We believe that machine learning will make a significant leap forward when it will incorporate external world knowledge and context into its decision-making processes. \textcolor{black}{For instance, an auto-encoder, if properly trained, can be very useful to compress the input and still capture relevant information from the data. But auto-encoders work in a purely data-driven (inductive) fashion. However, whenever the observation is the product of a language, meant in a very broad sense as a set of entities and logical rules to combine them, one can introduce the logic of the language inside the mechanisms of an auto-encoder, to drive its learning phase towards solutions that are coherent with the underlying language.}

6G networks can facilitate the merge between machine learning and knowledge representation systems. Semantic communication will in fact push for the introduction of knowledge base systems across the nodes of the network, to enable semantic interpretation. Within this semantically-enriched context, new machine learning algorithms can benefit from the inclusion of knowledge representation and reasoning schemes. At the same time, semantic  learning offers more capabilities to further optimize the use of network resources, focusing on semantic and goal-oriented aspects. \textcolor{black}{In this context, transfer learning and knowledge sharing mechanisms will play a key role in making semantic communication and machine learning algorithms more efficient and reliable.} 
\section{Conclusions}
\label{Conclusions}
In this paper, we have proposed a new vision of 6G wireless networks, where semantic and goal-oriented communications are the key actors of a paradigm shift, with respect to the common Shannon's framework, that has the potential of bringing enormous benefits. Increased effectiveness and reliability can be attained without necessarily increasing bandwidth or energy, but actually identifying the {\it relevant} information, i.e. the information strictly necessary to enable the receiver to extract the intended meaning correctly or to actuate the right procedures for achieving a predefined goal in the most efficient way. 
This approach capitalizes on the largely untapped capabilities of communication, computation and caching systems, on one side, and on knowledge representation tools on the other side, to distil the relevant information from the rest, selectively transmitting, processing, inferring and remembering only information relevant to goals defined by the interacting parties. 

The new philosophy breaks the usual trend aimed to provide more and more resources, like, e.g., energy or bandwidth, to enable more and more sophisticated services. This breakdown is at the core of a vision that looks at sustainability as the key property of future networks.

The challenge brought by the new approach is the implementation of distributed computing mechanisms able to learn and extract meaning from data, exploiting proper knowledge representation systems, and to identify and exploit the strictly relevant information in goal-oriented communications. In this new framework, learning can greatly benefit from the introduction of semantic aspects, in order to pass from a purely inductive strategy to the interplay of inductive and deductive mechanism, learning from examples, but also building abstract models guiding next learning and so, similarly to the way the human brain works. 

\section*{Acknowledgment}
This work has been partly funded by the European Commission through the H2020 project through the \mbox{RISE-6G}, \mbox{HEXA-X} (Grant Agreement no. 101015956) and DEDICAT-6G (Grant Agreement no. 101016499), H2020/ Taiwan Project \mbox{5G CONNI} (Grant Agreement no.861459) projects and, by MIUR (under the PRIN Liquid Edge contract).

\bibliographystyle{elsarticle-num-names}
\bibliography{sample}

\begin{thebibliography}{96}
\expandafter\ifx\csname natexlab\endcsname\relax\def\natexlab#1{#1}\fi
\providecommand{\url}[1]{\texttt{#1}}
\providecommand{\href}[2]{#2}
\providecommand{\path}[1]{#1}
\providecommand{\DOIprefix}{doi:}
\providecommand{\ArXivprefix}{arXiv:}
\providecommand{\URLprefix}{URL: }
\providecommand{\Pubmedprefix}{pmid:}
\providecommand{\doi}[1]{\href{http://dx.doi.org/#1}{\path{#1}}}
\providecommand{\Pubmed}[1]{\href{pmid:#1}{\path{#1}}}
\providecommand{\bibinfo}[2]{#2}
\ifx\xfnm\relax \def\xfnm[#1]{\unskip,\space#1}\fi
\bibitem[{Calvanese~Strinati et~al.(2019)Calvanese~Strinati, Barbarossa
  et~al.}]{Calvanese2019}
\bibinfo{author}{E.~Calvanese~Strinati}, \bibinfo{author}{S.~Barbarossa},
  et~al.,
\newblock \bibinfo{title}{{6{G}}: The next frontier: From holographic messaging
  to artificial intelligence using subterahertz and visible light
  communication},
\newblock \bibinfo{journal}{IEEE Vehicular Technology Magazine}
  \bibinfo{volume}{14} (\bibinfo{year}{2019}) \bibinfo{pages}{42--50}.
\bibitem[{Belot et~al.(2020)Belot, Gonz{\'a}lez~Jim{\'e}nez, Mercier, and
  Dor{\'e}}]{belot2020spectrum}
\bibinfo{author}{D.~Belot}, \bibinfo{author}{J.~L. Gonz{\'a}lez~Jim{\'e}nez},
  \bibinfo{author}{E.~Mercier}, \bibinfo{author}{J.-B. Dor{\'e}},
\newblock \bibinfo{title}{Spectrum above 90 {GH}z for wireless connectivity:
  Opportunities and challenges for 6{G}},
\newblock \bibinfo{journal}{Microwave Journal} \bibinfo{volume}{63}
  (\bibinfo{year}{2020}).
\bibitem[{Shannon(1948)}]{shannon1948mathematical}
\bibinfo{author}{C.~E. Shannon},
\newblock \bibinfo{title}{A mathematical theory of communication},
\newblock \bibinfo{journal}{The Bell system technical journal}
  \bibinfo{volume}{27} (\bibinfo{year}{1948}) \bibinfo{pages}{379--423}.
\bibitem[{Popovski et~al.(2020)Popovski, Simeone, Boccardi, G{\"u}nd{\"u}z, and
  Sahin}]{popovski2020semantic}
\bibinfo{author}{P.~Popovski}, \bibinfo{author}{O.~Simeone},
  \bibinfo{author}{F.~Boccardi}, \bibinfo{author}{D.~G{\"u}nd{\"u}z},
  \bibinfo{author}{O.~Sahin},
\newblock \bibinfo{title}{Semantic-effectiveness filtering and control for
  post-{5G} wireless connectivity},
\newblock \bibinfo{journal}{Journal of the Indian Institute of Science}
  \bibinfo{volume}{100} (\bibinfo{year}{2020}) \bibinfo{pages}{435--443}.
\bibitem[{Xie et~al.(2020)Xie, Ras, van Gerven, and Doran}]{xie2020explainable}
\bibinfo{author}{N.~Xie}, \bibinfo{author}{G.~Ras}, \bibinfo{author}{M.~van
  Gerven}, \bibinfo{author}{D.~Doran},
\newblock \bibinfo{title}{Explainable deep learning: A field guide for the
  uninitiated},
\newblock \bibinfo{journal}{arXiv preprint arXiv:2004.14545}
  (\bibinfo{year}{2020}).
\bibitem[{{Letaief} et~al.(2019){Letaief}, {Chen}, {Shi}, {Zhang}, and
  {Zhang}}]{Letaief20196GRoadMap}
\bibinfo{author}{K.~B. {Letaief}}, \bibinfo{author}{W.~{Chen}},
  \bibinfo{author}{Y.~{Shi}}, \bibinfo{author}{J.~{Zhang}},
  \bibinfo{author}{Y.~A. {Zhang}},
\newblock \bibinfo{title}{The roadmap to 6g: Ai empowered wireless networks},
\newblock \bibinfo{journal}{IEEE Communications Magazine} \bibinfo{volume}{57}
  (\bibinfo{year}{2019}) \bibinfo{pages}{84--90}.
  \DOIprefix\doi{10.1109/MCOM.2019.1900271}.
\bibitem[{Doré and et~al.(2020)}]{dore2020SubTHz}
\bibinfo{author}{J.-B. Doré}, \bibinfo{author}{et~al.},
\newblock \bibinfo{title}{Technology roadmap for beyond 5{G} wireless
  connectivity in {D}-band},
\newblock in: \bibinfo{booktitle}{2nd 6{G} Summit}, \bibinfo{year}{2020}.
\bibitem[{NET-2030(2019)}]{network203020196Gusecases}
\bibinfo{author}{F.~NET-2030},
\newblock \bibinfo{title}{A blueprint of technology, applications and market
  drivers towards the year 2030 and beyond},
\newblock \bibinfo{journal}{White Paper, Focus Group on Technologies for
  Network 2030 (FG NET-2030)}  (\bibinfo{year}{2019}).
\bibitem[{{Zhang} et~al.(2019){Zhang}, {Xiao}, {Ma}, {Xiao}, {Ding}, {Lei},
  {Karagiannidis}, and {Fan}}]{Zhang2019UseCase6G}
\bibinfo{author}{Z.~{Zhang}}, \bibinfo{author}{Y.~{Xiao}},
  \bibinfo{author}{Z.~{Ma}}, \bibinfo{author}{M.~{Xiao}},
  \bibinfo{author}{Z.~{Ding}}, \bibinfo{author}{X.~{Lei}},
  \bibinfo{author}{G.~K. {Karagiannidis}}, \bibinfo{author}{P.~{Fan}},
\newblock \bibinfo{title}{6g wireless networks: Vision, requirements,
  architecture, and key technologies},
\newblock \bibinfo{journal}{IEEE Vehicular Technology Magazine}
  \bibinfo{volume}{14} (\bibinfo{year}{2019}) \bibinfo{pages}{28--41}.
  \DOIprefix\doi{10.1109/MVT.2019.2921208}.
\bibitem[{{Yang} et~al.(2019){Yang}, {Xiao}, {Xiao}, and
  {Li}}]{Yang2019UseCase6G}
\bibinfo{author}{P.~{Yang}}, \bibinfo{author}{Y.~{Xiao}},
  \bibinfo{author}{M.~{Xiao}}, \bibinfo{author}{S.~{Li}},
\newblock \bibinfo{title}{6g wireless communications: Vision and potential
  techniques},
\newblock \bibinfo{journal}{IEEE Network} \bibinfo{volume}{33}
  (\bibinfo{year}{2019}) \bibinfo{pages}{70--75}.
  \DOIprefix\doi{10.1109/MNET.2019.1800418}.
\bibitem[{Nandana~Rajatheva(2020)}]{Nandana2020-6GBroadbandWhitepaper}
\bibinfo{author}{e.~a. Nandana~Rajatheva},
\newblock \bibinfo{title}{White paper on broadband connectivity in 6g},
\newblock \bibinfo{journal}{arXiv preprint arXiv:2004.14247v1}
  (\bibinfo{year}{2020}).
\bibitem[{Calvanese~Strinati et~al.(2020)Calvanese~Strinati, Barbarossa, Choi,
  Pietrabissa, Giuseppi, {De Santis}, Vidal, Becvar, Haustein, Nicolas,
  Costanzo, Kim, and Kim}]{Calvanese2020Sky6G}
\bibinfo{author}{E.~Calvanese~Strinati}, \bibinfo{author}{S.~Barbarossa},
  \bibinfo{author}{T.~Choi}, \bibinfo{author}{A.~Pietrabissa},
  \bibinfo{author}{A.~Giuseppi}, \bibinfo{author}{E.~{De Santis}},
  \bibinfo{author}{J.~Vidal}, \bibinfo{author}{Z.~Becvar},
  \bibinfo{author}{T.~Haustein}, \bibinfo{author}{C.~Nicolas},
  \bibinfo{author}{F.~Costanzo}, \bibinfo{author}{J.~Kim},
  \bibinfo{author}{I.~Kim},
\newblock \bibinfo{title}{6{G} in the sky: On-demand intelligence at the edge
  of {3D} networks},
\newblock \bibinfo{journal}{ETRI Journal}
  \bibinfo{volume}{10.4218/etrij.2020-0205} (\bibinfo{year}{2020}).
\bibitem[{Himesh et~al.(2018)Himesh, Gouda, Ramesh, V., Rakesh, Mohapatra, Rao,
  Sahoo, and Ajilesh}]{Himesh2018SmartAgriculture}
\bibinfo{author}{E.~V. S.~P. Himesh, S.and~Rao}, \bibinfo{author}{K.~C. Gouda},
  \bibinfo{author}{Ramesh}, \bibinfo{author}{K.~V.},
  \bibinfo{author}{V.~Rakesh}, \bibinfo{author}{G.~N. Mohapatra},
  \bibinfo{author}{B.~K. Rao}, \bibinfo{author}{S.~K. Sahoo},
  \bibinfo{author}{P.~Ajilesh},
\newblock \bibinfo{title}{Digital revolution and big data: A new revolution in
  agriculture},
\newblock \bibinfo{journal}{CAB Reviews} \bibinfo{volume}{27}
  (\bibinfo{year}{2018}) \bibinfo{pages}{1--7}.
  \DOIprefix\doi{10.1079/PAVSNNR201813021}.
\bibitem[{{Tariq} et~al.(2020){Tariq}, {Khandaker}, {Wong}, {Imran}, {Bennis},
  and {Debbah}}]{Tariq2020UseCase6G}
\bibinfo{author}{F.~{Tariq}}, \bibinfo{author}{M.~R.~A. {Khandaker}},
  \bibinfo{author}{K.~K. {Wong}}, \bibinfo{author}{M.~A. {Imran}},
  \bibinfo{author}{M.~{Bennis}}, \bibinfo{author}{M.~{Debbah}},
\newblock \bibinfo{title}{A speculative study on 6g},
\newblock \bibinfo{journal}{IEEE Wireless Communications} \bibinfo{volume}{27}
  (\bibinfo{year}{2020}) \bibinfo{pages}{118--125}.
  \DOIprefix\doi{10.1109/MWC.001.1900488}.
\bibitem[{ConsumerLab(2020)}]{Ericsson6GSenses}
\bibinfo{author}{E.~ConsumerLab},
\newblock \bibinfo{title}{Internet of senses report: Ericsson consumerlab hot
  consumer trends report.},
\newblock \bibinfo{journal}{Available online at
  https://www.ericsson.com/en/reports-and-papers/consumerlab/reports/10-hot-consumer-trends-2030}
   (\bibinfo{year}{2020}).
\bibitem[{David and Berndt(2018)}]{david20186g}
\bibinfo{author}{K.~David}, \bibinfo{author}{H.~Berndt},
\newblock \bibinfo{title}{6{G} vision and requirements: Is there any need for
  beyond 5{G}?},
\newblock \bibinfo{journal}{IEEE Vehicular Technology Magazine}
  \bibinfo{volume}{13} (\bibinfo{year}{2018}) \bibinfo{pages}{72--80}.
\bibitem[{Saad et~al.(2019)Saad, Bennis, and Chen}]{saad2019vision}
\bibinfo{author}{W.~Saad}, \bibinfo{author}{M.~Bennis},
  \bibinfo{author}{M.~Chen},
\newblock \bibinfo{title}{A vision of 6g wireless systems: Applications,
  trends, technologies, and open research problems},
\newblock \bibinfo{journal}{IEEE network} \bibinfo{volume}{34}
  (\bibinfo{year}{2019}) \bibinfo{pages}{134--142}.
\bibitem[{Shiroishi et~al.(2018)Shiroishi, Uchiyama, and
  Suzuki}]{shiroishi2018society}
\bibinfo{author}{Y.~Shiroishi}, \bibinfo{author}{K.~Uchiyama},
  \bibinfo{author}{N.~Suzuki},
\newblock \bibinfo{title}{Society 5.0: For human security and well-being},
\newblock \bibinfo{journal}{Computer} \bibinfo{volume}{51}
  (\bibinfo{year}{2018}) \bibinfo{pages}{91--95}.
\bibitem[{{Giordani} et~al.(2020){Giordani}, {Polese}, {Mezzavilla}, {Rangan},
  and {Zorzi}}]{Giordani2020}
\bibinfo{author}{M.~{Giordani}}, \bibinfo{author}{M.~{Polese}},
  \bibinfo{author}{M.~{Mezzavilla}}, \bibinfo{author}{S.~{Rangan}},
  \bibinfo{author}{M.~{Zorzi}},
\newblock \bibinfo{title}{Toward 6{G} networks: Use cases and technologies},
\newblock \bibinfo{journal}{IEEE Communications Magazine} \bibinfo{volume}{58}
  (\bibinfo{year}{2020}) \bibinfo{pages}{55--61}.
\bibitem[{Nakamura(2020)}]{nakamura20205g}
\bibinfo{author}{T.~Nakamura},
\newblock \bibinfo{title}{5{G} evolution and 6{G}},
\newblock in: \bibinfo{booktitle}{2020 International Symposium on VLSI Design,
  Automation and Test (VLSI-DAT)}, \bibinfo{organization}{IEEE},
  \bibinfo{year}{2020}, p.~\bibinfo{pages}{1}.
\bibitem[{Maier and Ebrahimzadeh(2020)}]{maier2020toward}
\bibinfo{author}{M.~Maier}, \bibinfo{author}{A.~Ebrahimzadeh},
  \bibinfo{title}{Toward 6G: A New Era of Convergence},
  \bibinfo{publisher}{John Wiley \& Sons}, \bibinfo{year}{2020}.
\bibitem[{{Tariq} et~al.(2020){Tariq}, {Khandaker}, {Wong}, {Imran}, {Bennis},
  and {Debbah}}]{Tariq2020}
\bibinfo{author}{F.~{Tariq}}, \bibinfo{author}{M.~R.~A. {Khandaker}},
  \bibinfo{author}{K.~{Wong}}, \bibinfo{author}{M.~A. {Imran}},
  \bibinfo{author}{M.~{Bennis}}, \bibinfo{author}{M.~{Debbah}},
\newblock \bibinfo{title}{A speculative study on 6{G}},
\newblock \bibinfo{journal}{IEEE Wireless Communications} \bibinfo{volume}{27}
  (\bibinfo{year}{2020}) \bibinfo{pages}{118--125}.
\bibitem[{Li(2018)}]{li2018towards}
\bibinfo{author}{R.~Li},
\newblock \bibinfo{title}{Towards a new internet for the year 2030 and beyond},
\newblock in: \bibinfo{booktitle}{Proc. 3rd Annu. ITU IMT-2020/5G Workshop Demo
  Day}, \bibinfo{year}{2018}, pp. \bibinfo{pages}{1--21}.
\bibitem[{Berardinelli et~al.(2018)Berardinelli, Mahmood, Rodriguez, and
  Mogensen}]{berardinelli2018beyond}
\bibinfo{author}{G.~Berardinelli}, \bibinfo{author}{N.~H. Mahmood},
  \bibinfo{author}{I.~Rodriguez}, \bibinfo{author}{P.~Mogensen},
\newblock \bibinfo{title}{Beyond 5{G} wireless irt for industry 4.0: Design
  principles and spectrum aspects},
\newblock in: \bibinfo{booktitle}{2018 IEEE Globecom Workshops (GC Wkshps)},
  \bibinfo{organization}{IEEE}, \bibinfo{year}{2018}, pp.
  \bibinfo{pages}{1--6}.
\bibitem[{Strinati and et~al.(2020)}]{calvanese2020IIoT}
\bibinfo{author}{E.~C. Strinati}, \bibinfo{author}{et~al.},
\newblock \bibinfo{title}{Beyond 5{G} private networks: the 5{G} conni
  perspective},
\newblock in: \bibinfo{booktitle}{2020 IEEE Globecom Workshops (FIIoT
  workshop)}, \bibinfo{organization}{IEEE}, \bibinfo{year}{2020}.
\bibitem[{Vora et~al.(2020)Vora, Karthik, and Saravanan}]{vora2020provisioning}
\bibinfo{author}{R.~Vora}, \bibinfo{author}{R.~Karthik},
  \bibinfo{author}{M.~Saravanan},
\newblock \bibinfo{title}{Provisioning of broadband communication for
  passengers in hyperloop using 5{G} networks},
\newblock in: \bibinfo{booktitle}{ICT for Competitive Strategies: Proceedings
  of 4th International Conference on Information and Communication Technology
  for Competitive Strategies (ICTCS 2019), December 13th-14th, 2019, Udaipur,
  India}, \bibinfo{organization}{CRC Press}, \bibinfo{year}{2020}, p.
  \bibinfo{pages}{317}.
\bibitem[{itu(2015)}]{itu2015expraffic}
\bibinfo{title}{Tmt traffic estimates for the years 2020 to 2030},
\newblock in: \bibinfo{booktitle}{Proc. M Series Mobile, radiodetermination,
  amateur and related satellite services. Report ITU-R M.2370-0},
  \bibinfo{year}{2015}.
\bibitem[{et~al.(2020)}]{haas2020Lifi6G}
\bibinfo{author}{H.~H. et~al.},
\newblock \bibinfo{title}{Introduction to indoor networking concepts and
  challenges in lifi},
\newblock \bibinfo{journal}{IEEE/OSA Journal of Optical Communications and
  Networking} \bibinfo{volume}{12} (\bibinfo{year}{2020})
  \bibinfo{pages}{190--203}.
\bibitem[{Oueis and Strinati(2016)}]{oueis2016uplink}
\bibinfo{author}{J.~Oueis}, \bibinfo{author}{E.~C. Strinati},
\newblock \bibinfo{title}{Uplink traffic in future mobile networks: Pulling the
  alarm},
\newblock in: \bibinfo{booktitle}{International Conference on Cognitive Radio
  Oriented Wireless Networks}, \bibinfo{organization}{Springer},
  \bibinfo{year}{2016}, pp. \bibinfo{pages}{583--593}.
\bibitem[{Solutions(2013)}]{solutions2013networks}
\bibinfo{author}{N.~Solutions}, \bibinfo{title}{Networks,“{N}okia solutions
  and networks td-lte frame configuration primer,”}, \bibinfo{year}{2013}.
\bibitem[{Markus et~al.(2009)Markus, Oliver, Dieter, Dietrich, Ali, and
  Calvanese~Strinati}]{earth2009}
\bibinfo{author}{G.~Markus}, \bibinfo{author}{B.~Oliver},
  \bibinfo{author}{F.~Dieter}, \bibinfo{author}{Z.~Dietrich},
  \bibinfo{author}{I.~M. Ali}, \bibinfo{author}{E.~Calvanese~Strinati},
\newblock \bibinfo{title}{Earth—energy aware radio and network technologies},
\newblock in: \bibinfo{booktitle}{2009 IEEE 20th International Symposium on
  Personal, Indoor and Mobile Radio Communications},
  \bibinfo{organization}{IEEE}, \bibinfo{year}{2009}, pp.
  \bibinfo{pages}{1--5}.
\bibitem[{Liu et~al.(2016)Liu, Wu, Xia, Zhao, Chen, Yang, and
  Hanzo}]{IEEEVTM-Liu2016}
\bibinfo{author}{Liu}, \bibinfo{author}{J.~Wu}, \bibinfo{author}{P.~Xia},
  \bibinfo{author}{S.~Zhao}, \bibinfo{author}{W.~Chen},
  \bibinfo{author}{Y.~Yang}, \bibinfo{author}{L.~Hanzo},
\newblock \bibinfo{title}{Charging unplugged: Will distributed laser charging
  for mobile wireless power transfer work?},
\newblock \bibinfo{journal}{IEEE Vehicular Technology Magazine}
  \bibinfo{volume}{11} (\bibinfo{year}{2016}) \bibinfo{pages}{36--45}.
\bibitem[{{Yaacoub} and {Alouini}(2020)}]{Yaacoub6GRuralAreas}
\bibinfo{author}{E.~{Yaacoub}}, \bibinfo{author}{M.~{Alouini}},
\newblock \bibinfo{title}{A key 6g challenge and opportunity—connecting the
  base of the pyramid: A survey on rural connectivity},
\newblock \bibinfo{journal}{Proceedings of the IEEE} \bibinfo{volume}{108}
  (\bibinfo{year}{2020}) \bibinfo{pages}{533--582}.
  \DOIprefix\doi{10.1109/JPROC.2020.2976703}.
\bibitem[{Berners-Lee et~al.(2001)Berners-Lee, Hendler, and
  Lassila.}]{Berners2001SemanticWeb}
\bibinfo{author}{T.~Berners-Lee}, \bibinfo{author}{J.~Hendler},
  \bibinfo{author}{O.~Lassila.},
\newblock \bibinfo{title}{The semantic web},
\newblock \bibinfo{journal}{Scientific American}  (\bibinfo{year}{2001})
  \bibinfo{pages}{34--43}.
\bibitem[{Weaver(1953)}]{weaver1953recent}
\bibinfo{author}{W.~Weaver},
\newblock \bibinfo{title}{Recent contributions to the mathematical theory of
  communication},
\newblock \bibinfo{journal}{ETC: a review of general semantics}
  (\bibinfo{year}{1953}) \bibinfo{pages}{261--281}.
\bibitem[{Floridi(2002)}]{floridi2002philosophy}
\bibinfo{author}{L.~Floridi},
\newblock \bibinfo{title}{What is the philosophy of information?},
\newblock \bibinfo{journal}{Metaphilosophy} \bibinfo{volume}{33}
  (\bibinfo{year}{2002}) \bibinfo{pages}{123--145}.
\bibitem[{Devlin(1995)}]{devlin1995logic}
\bibinfo{author}{K.~Devlin}, \bibinfo{title}{Logic and information},
  \bibinfo{publisher}{Cambridge University Press}, \bibinfo{year}{1995}.
\bibitem[{Kohlas(2012)}]{kohlas2012information}
\bibinfo{author}{J.~Kohlas}, \bibinfo{title}{Information algebras: Generic
  structures for inference}, \bibinfo{publisher}{Springer Science \& Business
  Media}, \bibinfo{year}{2012}.
\bibitem[{Barwise et~al.(1997)Barwise, Seligman
  et~al.}]{barwise1997information}
\bibinfo{author}{J.~Barwise}, \bibinfo{author}{J.~Seligman}, et~al.,
  \bibinfo{title}{Information flow: the logic of distributed systems},
  \bibinfo{publisher}{Cambridge University Press}, \bibinfo{year}{1997}.
\bibitem[{Nielsen and Chuang(2002)}]{nielsen2002quantum}
\bibinfo{author}{M.~A. Nielsen}, \bibinfo{author}{I.~Chuang},
  \bibinfo{title}{Quantum computation and quantum information},
  \bibinfo{year}{2002}.
\bibitem[{Calude(2013)}]{calude2013information}
\bibinfo{author}{C.~S. Calude}, \bibinfo{title}{Information and randomness: an
  algorithmic perspective}, \bibinfo{publisher}{Springer Science \& Business
  Media}, \bibinfo{year}{2013}.
\bibitem[{Chaitin(1977)}]{chaitin1977algorithmic}
\bibinfo{author}{G.~J. Chaitin},
\newblock \bibinfo{title}{Algorithmic information theory},
\newblock \bibinfo{journal}{IBM journal of research and development}
  \bibinfo{volume}{21} (\bibinfo{year}{1977}) \bibinfo{pages}{350--359}.
\bibitem[{Willems and Kalker(2005)}]{willems2005semantic}
\bibinfo{author}{F.~M. Willems}, \bibinfo{author}{T.~Kalker},
\newblock \bibinfo{title}{Semantic compaction, transmission, and compression
  codes},
\newblock in: \bibinfo{booktitle}{Proceedings. International Symposium on
  Information Theory, 2005. ISIT 2005.}, \bibinfo{organization}{IEEE},
  \bibinfo{year}{2005}, pp. \bibinfo{pages}{214--218}.
\bibitem[{Juba(2011)}]{juba2011universal}
\bibinfo{author}{B.~Juba}, \bibinfo{title}{Universal semantic communication},
  \bibinfo{publisher}{Springer Science \& Business Media},
  \bibinfo{year}{2011}.
\bibitem[{Bao et~al.(2011)Bao, Basu, Dean, Partridge, Swami, Leland, and
  Hendler}]{bao2011towards}
\bibinfo{author}{J.~Bao}, \bibinfo{author}{P.~Basu}, \bibinfo{author}{M.~Dean},
  \bibinfo{author}{C.~Partridge}, \bibinfo{author}{A.~Swami},
  \bibinfo{author}{W.~Leland}, \bibinfo{author}{J.~A. Hendler},
\newblock \bibinfo{title}{Towards a theory of semantic communication},
\newblock in: \bibinfo{booktitle}{2011 IEEE Network Science Workshop},
  \bibinfo{organization}{IEEE}, \bibinfo{year}{2011}, pp.
  \bibinfo{pages}{110--117}.
\bibitem[{Goldreich et~al.(2012)Goldreich, Juba, and
  Sudan}]{goldreich2012theory}
\bibinfo{author}{O.~Goldreich}, \bibinfo{author}{B.~Juba},
  \bibinfo{author}{M.~Sudan},
\newblock \bibinfo{title}{A theory of goal-oriented communication},
\newblock \bibinfo{journal}{Journal of the ACM (JACM)} \bibinfo{volume}{59}
  (\bibinfo{year}{2012}) \bibinfo{pages}{1--65}.
\bibitem[{Basu et~al.(2014)Basu, Bao, Dean, and Hendler}]{basu2014preserving}
\bibinfo{author}{P.~Basu}, \bibinfo{author}{J.~Bao}, \bibinfo{author}{M.~Dean},
  \bibinfo{author}{J.~Hendler},
\newblock \bibinfo{title}{Preserving quality of information by using semantic
  relationships},
\newblock \bibinfo{journal}{Pervasive and Mobile Computing}
  \bibinfo{volume}{11} (\bibinfo{year}{2014}) \bibinfo{pages}{188--202}.
\bibitem[{G{\"u}ler et~al.(2018)G{\"u}ler, Yener, and
  Swami}]{guler2018semantic}
\bibinfo{author}{B.~G{\"u}ler}, \bibinfo{author}{A.~Yener},
  \bibinfo{author}{A.~Swami},
\newblock \bibinfo{title}{The semantic communication game},
\newblock \bibinfo{journal}{IEEE Transactions on Cognitive Communications and
  Networking} \bibinfo{volume}{4} (\bibinfo{year}{2018})
  \bibinfo{pages}{787--802}.
\bibitem[{Kountouris and Pappas(2020)}]{kountouris2020semantics}
\bibinfo{author}{M.~Kountouris}, \bibinfo{author}{N.~Pappas},
\newblock \bibinfo{title}{Semantics-empowered communication for networked
  intelligent systems},
\newblock \bibinfo{journal}{arXiv preprint arXiv:2007.11579}
  (\bibinfo{year}{2020}).
\bibitem[{Xie et~al.(2020)Xie, Qin, Li, and Juang}]{xie2020deep}
\bibinfo{author}{H.~Xie}, \bibinfo{author}{Z.~Qin}, \bibinfo{author}{G.~Y. Li},
  \bibinfo{author}{B.-H. Juang},
\newblock \bibinfo{title}{Deep learning enabled semantic communication
  systems},
\newblock \bibinfo{journal}{arXiv preprint arXiv:2006.10685}
  (\bibinfo{year}{2020}).
\bibitem[{Dretske(1981)}]{dretske1981knowledge}
\bibinfo{author}{F.~Dretske},
\newblock \bibinfo{title}{Knowledge and the flow of information}
  (\bibinfo{year}{1981}).
\bibitem[{Russell and Norvig(2020)}]{russell2010artificial}
\bibinfo{author}{S.~J. Russell}, \bibinfo{author}{P.~Norvig},
  \bibinfo{title}{Artificial intelligence-a modern approach (4th edition)},
  \bibinfo{year}{2020}.
\bibitem[{Chein and Mugnier(2008)}]{chein2008graph}
\bibinfo{author}{M.~Chein}, \bibinfo{author}{M.-L. Mugnier},
  \bibinfo{title}{Graph-based knowledge representation: computational
  foundations of conceptual graphs}, \bibinfo{publisher}{Springer Science \&
  Business Media}, \bibinfo{year}{2008}.
\bibitem[{Farsad et~al.(2018)Farsad, Rao, and Goldsmith}]{farsad2018deep}
\bibinfo{author}{N.~Farsad}, \bibinfo{author}{M.~Rao},
  \bibinfo{author}{A.~Goldsmith},
\newblock \bibinfo{title}{Deep learning for joint source-channel coding of
  text},
\newblock in: \bibinfo{booktitle}{2018 IEEE International Conference on
  Acoustics, Speech and Signal Processing (ICASSP)},
  \bibinfo{organization}{IEEE}, \bibinfo{year}{2018}, pp.
  \bibinfo{pages}{2326--2330}.
\bibitem[{Choi and Baji{\'c}(2019)}]{choi2019deep}
\bibinfo{author}{H.~Choi}, \bibinfo{author}{I.~V. Baji{\'c}},
\newblock \bibinfo{title}{Deep frame prediction for video coding},
\newblock \bibinfo{journal}{IEEE Transactions on Circuits and Systems for Video
  Technology}  (\bibinfo{year}{2019}).
\bibitem[{Hohwy(2013)}]{hohwy2013predictive}
\bibinfo{author}{J.~Hohwy}, \bibinfo{title}{The predictive mind},
  \bibinfo{publisher}{Oxford University Press}, \bibinfo{year}{2013}.
\bibitem[{Clark(2015)}]{clark2015surfing}
\bibinfo{author}{A.~Clark}, \bibinfo{title}{Surfing uncertainty: Prediction,
  action, and the embodied mind}, \bibinfo{publisher}{Oxford University Press},
  \bibinfo{year}{2015}.
\bibitem[{Clark(2013)}]{clark2013whatever}
\bibinfo{author}{A.~Clark},
\newblock \bibinfo{title}{Whatever next? predictive brains, situated agents,
  and the future of cognitive science},
\newblock \bibinfo{journal}{Behavioral and brain sciences} \bibinfo{volume}{36}
  (\bibinfo{year}{2013}) \bibinfo{pages}{181--204}.
\bibitem[{Juba and Sudan(2008)}]{juba2008universal}
\bibinfo{author}{B.~Juba}, \bibinfo{author}{M.~Sudan},
\newblock \bibinfo{title}{Universal semantic communication ii: A theory of
  goal-oriented communication},
\newblock in: \bibinfo{booktitle}{Electronic Colloquium on Computational
  Complexity (ECCC)}, volume~\bibinfo{volume}{15}, \bibinfo{year}{2008}.
\bibitem[{Park et~al.(2019)Park, Samarakoon, Bennis, and
  Debbah}]{park2018wireless}
\bibinfo{author}{J.~Park}, \bibinfo{author}{S.~Samarakoon},
  \bibinfo{author}{M.~Bennis}, \bibinfo{author}{M.~Debbah},
\newblock \bibinfo{title}{Wireless network intelligence at the edge},
\newblock \bibinfo{journal}{Proceedings of the IEEE} \bibinfo{volume}{107}
  (\bibinfo{year}{2019}) \bibinfo{pages}{2204--2239}.
\bibitem[{{Skatchkovsky} and {Simeone}(2019)}]{Skatchkovsky2019}
\bibinfo{author}{N.~{Skatchkovsky}}, \bibinfo{author}{O.~{Simeone}},
\newblock \bibinfo{title}{Optimizing pipelined computation and communication
  for latency-constrained edge learning},
\newblock \bibinfo{journal}{IEEE Communications Letters} \bibinfo{volume}{23}
  (\bibinfo{year}{2019}) \bibinfo{pages}{1542--1546}.
\bibitem[{Mohammad and Sorour(2019)}]{Mohammad2019}
\bibinfo{author}{U.~Mohammad}, \bibinfo{author}{S.~Sorour},
\newblock \bibinfo{title}{Adaptive task allocation for mobile edge learning},
\newblock \bibinfo{journal}{Available online at
  https://arxiv.org/abs/1811.03748}  (\bibinfo{year}{2019}).
\bibitem[{Amiri and Gunduz(2019)}]{Amir2019}
\bibinfo{author}{M.~M. Amiri}, \bibinfo{author}{D.~Gunduz},
\newblock \bibinfo{title}{Machine learning at the wireless edge: Distributed
  stochastic gradient descent over-the-air},
\newblock \bibinfo{journal}{Available online at
  https://arxiv.org/abs/1901.00844}  (\bibinfo{year}{2019}).
\bibitem[{Cover(1999)}]{cover1999elements}
\bibinfo{author}{T.~M. Cover}, \bibinfo{title}{Elements of information theory},
  \bibinfo{publisher}{John Wiley \& Sons}, \bibinfo{year}{1999}.
\bibitem[{Tishby et~al.(2000)Tishby, Pereira, and
  Bialek}]{tishby2000information}
\bibinfo{author}{N.~Tishby}, \bibinfo{author}{F.~C. Pereira},
  \bibinfo{author}{W.~Bialek},
\newblock \bibinfo{title}{The information bottleneck method},
\newblock \bibinfo{journal}{arXiv preprint physics/0004057}
  (\bibinfo{year}{2000}).
\bibitem[{Shamir et~al.(2010)Shamir, Sabato, and Tishby}]{shamir2010learning}
\bibinfo{author}{O.~Shamir}, \bibinfo{author}{S.~Sabato},
  \bibinfo{author}{N.~Tishby},
\newblock \bibinfo{title}{Learning and generalization with the information
  bottleneck},
\newblock \bibinfo{journal}{Theoretical Computer Science} \bibinfo{volume}{411}
  (\bibinfo{year}{2010}) \bibinfo{pages}{2696--2711}.
\bibitem[{{Lecun} et~al.(1998){Lecun}, {Bottou}, {Bengio}, and
  {Haffner}}]{726791}
\bibinfo{author}{Y.~{Lecun}}, \bibinfo{author}{L.~{Bottou}},
  \bibinfo{author}{Y.~{Bengio}}, \bibinfo{author}{P.~{Haffner}},
\newblock \bibinfo{title}{Gradient-based learning applied to document
  recognition},
\newblock \bibinfo{journal}{Proceedings of the IEEE} \bibinfo{volume}{86}
  (\bibinfo{year}{1998}) \bibinfo{pages}{2278--2324}.
\bibitem[{{Helwig} et~al.(2015){Helwig}, {Pignanelli}, and
  {Schütze}}]{7151267}
\bibinfo{author}{N.~{Helwig}}, \bibinfo{author}{E.~{Pignanelli}},
  \bibinfo{author}{A.~{Schütze}},
\newblock \bibinfo{title}{Condition monitoring of a complex hydraulic system
  using multivariate statistics},
\newblock in: \bibinfo{booktitle}{2015 IEEE International Instrumentation and
  Measurement Technology Conference (I2MTC) Proceedings}, \bibinfo{year}{2015},
  pp. \bibinfo{pages}{210--215}.
\bibitem[{Sch{\"o}lkopf et~al.(2000)Sch{\"o}lkopf, Smola, Williamson, and
  Bartlett}]{scholkopf2000new}
\bibinfo{author}{B.~Sch{\"o}lkopf}, \bibinfo{author}{A.~J. Smola},
  \bibinfo{author}{R.~C. Williamson}, \bibinfo{author}{P.~L. Bartlett},
\newblock \bibinfo{title}{New support vector algorithms},
\newblock \bibinfo{journal}{Neural computation} \bibinfo{volume}{12}
  (\bibinfo{year}{2000}) \bibinfo{pages}{1207--1245}.
  \DOIprefix\doi{10.1162/089976600300015565}.
\bibitem[{Goldfeld et~al.(2018)Goldfeld, Berg, Greenewald, Melnyk, Nguyen,
  Kingsbury, and Polyanskiy}]{goldfeld2018estimating}
\bibinfo{author}{Z.~Goldfeld}, \bibinfo{author}{E.~v.~d. Berg},
  \bibinfo{author}{K.~Greenewald}, \bibinfo{author}{I.~Melnyk},
  \bibinfo{author}{N.~Nguyen}, \bibinfo{author}{B.~Kingsbury},
  \bibinfo{author}{Y.~Polyanskiy},
\newblock \bibinfo{title}{Estimating information flow in deep neural networks},
\newblock \bibinfo{journal}{arXiv preprint arXiv:1810.05728}
  (\bibinfo{year}{2018}).
\bibitem[{Zhang et~al.(2018)Zhang, Cao, Jiang, Zhang, Li, and
  Yao}]{zhang2018ffs}
\bibinfo{author}{C.~Zhang}, \bibinfo{author}{Q.~Cao},
  \bibinfo{author}{H.~Jiang}, \bibinfo{author}{W.~Zhang},
  \bibinfo{author}{J.~Li}, \bibinfo{author}{J.~Yao},
\newblock \bibinfo{title}{Ffs-va: A fast filtering system for large-scale video
  analytics},
\newblock in: \bibinfo{booktitle}{Proceedings of the 47th International
  Conference on Parallel Processing}, \bibinfo{year}{2018}, pp.
  \bibinfo{pages}{1--10}.
\bibitem[{Zhou et~al.(2019)Zhou, Chen, Li, Zeng, Luo, and Zhang}]{zhou2019edge}
\bibinfo{author}{Z.~Zhou}, \bibinfo{author}{X.~Chen}, \bibinfo{author}{E.~Li},
  \bibinfo{author}{L.~Zeng}, \bibinfo{author}{K.~Luo},
  \bibinfo{author}{J.~Zhang},
\newblock \bibinfo{title}{Edge intelligence: Paving the last mile of artificial
  intelligence with edge computing},
\newblock \bibinfo{journal}{Proceedings of the IEEE} \bibinfo{volume}{107}
  (\bibinfo{year}{2019}) \bibinfo{pages}{1738--1762}.
\bibitem[{Peltonen et~al.(2020)Peltonen, Bennis, Capobianco, Debbah, Ding,
  Gil-Casti{\~n}eira, Jurmu, Karvonen, Kelanti, Kliks et~al.}]{peltonen20206g}
\bibinfo{author}{E.~Peltonen}, \bibinfo{author}{M.~Bennis},
  \bibinfo{author}{M.~Capobianco}, \bibinfo{author}{M.~Debbah},
  \bibinfo{author}{A.~Ding}, \bibinfo{author}{F.~Gil-Casti{\~n}eira},
  \bibinfo{author}{M.~Jurmu}, \bibinfo{author}{T.~Karvonen},
  \bibinfo{author}{M.~Kelanti}, \bibinfo{author}{A.~Kliks}, et~al.,
\newblock \bibinfo{title}{6{G} white paper on edge intelligence},
\newblock \bibinfo{journal}{arXiv preprint arXiv:2004.14850}
  (\bibinfo{year}{2020}).
\bibitem[{Letaief et~al.(2019)Letaief, Chen, Shi, Zhang, and
  Zhang}]{letaief2019roadmap}
\bibinfo{author}{K.~B. Letaief}, \bibinfo{author}{W.~Chen},
  \bibinfo{author}{Y.~Shi}, \bibinfo{author}{J.~Zhang},
  \bibinfo{author}{Y.-J.~A. Zhang},
\newblock \bibinfo{title}{The roadmap to 6{G}: Ai empowered wireless networks},
\newblock \bibinfo{journal}{IEEE Communications Magazine} \bibinfo{volume}{57}
  (\bibinfo{year}{2019}) \bibinfo{pages}{84--90}.
\bibitem[{Barbarossa et~al.(2018)Barbarossa, Sardellitti, Ceci, and
  Merluzzi}]{barbarossa2018edge}
\bibinfo{author}{S.~Barbarossa}, \bibinfo{author}{S.~Sardellitti},
  \bibinfo{author}{E.~Ceci}, \bibinfo{author}{M.~Merluzzi},
\newblock \bibinfo{title}{The edge cloud: A holistic view of communication,
  computation, and caching},
\newblock in: \bibinfo{booktitle}{Cooperative and Graph Signal Processing},
  \bibinfo{publisher}{Elsevier}, \bibinfo{year}{2018}, pp.
  \bibinfo{pages}{419--444}.
\bibitem[{{Ndikumana} et~al.(2020){Ndikumana}, {Tran}, {Ho}, {Han}, {Saad},
  {Niyato}, and {Hong}}]{Ndikumana2020ieeeTMCC4}
\bibinfo{author}{A.~{Ndikumana}}, \bibinfo{author}{N.~H. {Tran}},
  \bibinfo{author}{T.~M. {Ho}}, \bibinfo{author}{Z.~{Han}},
  \bibinfo{author}{W.~{Saad}}, \bibinfo{author}{D.~{Niyato}},
  \bibinfo{author}{C.~S. {Hong}},
\newblock \bibinfo{title}{Joint communication, computation, caching, and
  control in big data multi-access edge computing},
\newblock \bibinfo{journal}{IEEE Transactions on Mobile Computing}
  \bibinfo{volume}{19} (\bibinfo{year}{2020}) \bibinfo{pages}{1359--1374}.
  \DOIprefix\doi{10.1109/TMC.2019.2908403}.
\bibitem[{Anselme(2019)}]{Ndikumana2019C4PhD}
\bibinfo{author}{N.~Anselme}, \bibinfo{title}{IntelligentEdge: Joint
  Communication, Computation, Caching, and Control in Collaborative
  Multi-access Edge Computing}, Ph.D. thesis, Kyung Hee University, South
  Korea, \bibinfo{year}{2019}.
\bibitem[{{Wang} et~al.(2019){Wang}, {Gao}, {Fang}, {Sun}, and
  {Si}}]{Wang2019INFOCOM-C4}
\bibinfo{author}{Z.~{Wang}}, \bibinfo{author}{Y.~{Gao}},
  \bibinfo{author}{C.~{Fang}}, \bibinfo{author}{Y.~{Sun}},
  \bibinfo{author}{P.~{Si}},
\newblock \bibinfo{title}{Optimal control design for connected cruise control
  with edge computing, caching, and control},
\newblock in: \bibinfo{booktitle}{IEEE INFOCOM 2019 - IEEE Conference on
  Computer Communications Workshops (INFOCOM WKSHPS)}, \bibinfo{year}{2019},
  pp. \bibinfo{pages}{1--6}.
  \DOIprefix\doi{10.1109/INFOCOMWKSHPS47286.2019.9093766}.
\bibitem[{Sardellitti et~al.(2015)Sardellitti, Scutari, and
  Barbarossa}]{sardellitti2015joint}
\bibinfo{author}{S.~Sardellitti}, \bibinfo{author}{G.~Scutari},
  \bibinfo{author}{S.~Barbarossa},
\newblock \bibinfo{title}{Joint optimization of radio and computational
  resources for multicell mobile-edge computing},
\newblock \bibinfo{journal}{IEEE Transactions on Signal and Information
  Processing over Networks} \bibinfo{volume}{1} (\bibinfo{year}{2015})
  \bibinfo{pages}{89--103}.
\bibitem[{Mao et~al.(2017)Mao, Zhang, Song, and Letaief}]{mao2017stochastic}
\bibinfo{author}{Y.~Mao}, \bibinfo{author}{J.~Zhang},
  \bibinfo{author}{S.~Song}, \bibinfo{author}{K.~B. Letaief},
\newblock \bibinfo{title}{Stochastic joint radio and computational resource
  management for multi-user mobile-edge computing systems},
\newblock \bibinfo{journal}{IEEE Transactions on Wireless Communications}
  \bibinfo{volume}{16} (\bibinfo{year}{2017}) \bibinfo{pages}{5994--6009}.
\bibitem[{Merluzzi et~al.(2020)Merluzzi, Di~Lorenzo, Barbarossa, and
  Frascolla}]{merluzzi2020dynamic}
\bibinfo{author}{M.~Merluzzi}, \bibinfo{author}{P.~Di~Lorenzo},
  \bibinfo{author}{S.~Barbarossa}, \bibinfo{author}{V.~Frascolla},
\newblock \bibinfo{title}{Dynamic computation offloading in multi-access edge
  computing via ultra-reliable and low-latency communications},
\newblock \bibinfo{journal}{IEEE Transactions on Signal and Information
  Processing over Networks} \bibinfo{volume}{6} (\bibinfo{year}{2020})
  \bibinfo{pages}{342--356}.
\bibitem[{Chen et~al.(2019)Chen, Barbarossa, Wang, Giannakis, and
  Zhang}]{chen2019learning}
\bibinfo{author}{T.~Chen}, \bibinfo{author}{S.~Barbarossa},
  \bibinfo{author}{X.~Wang}, \bibinfo{author}{G.~B. Giannakis},
  \bibinfo{author}{Z.-L. Zhang},
\newblock \bibinfo{title}{Learning and management for internet of things:
  Accounting for adaptivity and scalability},
\newblock \bibinfo{journal}{Proceedings of the IEEE} \bibinfo{volume}{107}
  (\bibinfo{year}{2019}) \bibinfo{pages}{778--796}.
\bibitem[{LeCun et~al.(2015)LeCun, Bengio, and Hinton}]{lecun2015deep}
\bibinfo{author}{Y.~LeCun}, \bibinfo{author}{Y.~Bengio},
  \bibinfo{author}{G.~Hinton},
\newblock \bibinfo{title}{Deep learning},
\newblock \bibinfo{journal}{nature} \bibinfo{volume}{521}
  (\bibinfo{year}{2015}) \bibinfo{pages}{436--444}.
\bibitem[{Bronstein et~al.(2017)Bronstein, Bruna, LeCun, Szlam, and
  Vandergheynst}]{bronstein2017geometric}
\bibinfo{author}{M.~M. Bronstein}, \bibinfo{author}{J.~Bruna},
  \bibinfo{author}{Y.~LeCun}, \bibinfo{author}{A.~Szlam},
  \bibinfo{author}{P.~Vandergheynst},
\newblock \bibinfo{title}{Geometric deep learning: going beyond euclidean
  data},
\newblock \bibinfo{journal}{IEEE Signal Processing Magazine}
  \bibinfo{volume}{34} (\bibinfo{year}{2017}) \bibinfo{pages}{18--42}.
\bibitem[{Barbarossa and Sardellitti(2020)}]{barbarossa2020topological}
\bibinfo{author}{S.~Barbarossa}, \bibinfo{author}{S.~Sardellitti},
\newblock \bibinfo{title}{Topological signal processing over simplicial
  complexes},
\newblock \bibinfo{journal}{IEEE Transactions on Signal Processing}
  \bibinfo{volume}{68} (\bibinfo{year}{2020}) \bibinfo{pages}{2992--3007}.
\bibitem[{D{\"o}rner et~al.(2017)D{\"o}rner, Cammerer, Hoydis, and
  Ten~Brink}]{dorner2017deep}
\bibinfo{author}{S.~D{\"o}rner}, \bibinfo{author}{S.~Cammerer},
  \bibinfo{author}{J.~Hoydis}, \bibinfo{author}{S.~Ten~Brink},
\newblock \bibinfo{title}{Deep learning based communication over the air},
\newblock \bibinfo{journal}{IEEE Journal of Selected Topics in Signal
  Processing} \bibinfo{volume}{12} (\bibinfo{year}{2017})
  \bibinfo{pages}{132--143}.
\bibitem[{Balevi and Andrews(2019)}]{balevi2019one}
\bibinfo{author}{E.~Balevi}, \bibinfo{author}{J.~G. Andrews},
\newblock \bibinfo{title}{One-bit ofdm receivers via deep learning},
\newblock \bibinfo{journal}{IEEE Transactions on Communications}
  \bibinfo{volume}{67} (\bibinfo{year}{2019}) \bibinfo{pages}{4326--4336}.
\bibitem[{Farsad and Goldsmith(2018)}]{farsad2018neural}
\bibinfo{author}{N.~Farsad}, \bibinfo{author}{A.~Goldsmith},
\newblock \bibinfo{title}{Neural network detection of data sequences in
  communication systems},
\newblock \bibinfo{journal}{IEEE Transactions on Signal Processing}
  \bibinfo{volume}{66} (\bibinfo{year}{2018}) \bibinfo{pages}{5663--5678}.
\bibitem[{Ye et~al.(2018)Ye, Li, Juang, and Sivanesan}]{ye2018channel}
\bibinfo{author}{H.~Ye}, \bibinfo{author}{G.~Y. Li}, \bibinfo{author}{B.-H.~F.
  Juang}, \bibinfo{author}{K.~Sivanesan},
\newblock \bibinfo{title}{Channel agnostic end-to-end learning based
  communication systems with conditional gan},
\newblock in: \bibinfo{booktitle}{2018 IEEE Globecom Workshops (GC Wkshps)},
  \bibinfo{organization}{IEEE}, \bibinfo{year}{2018}, pp.
  \bibinfo{pages}{1--5}.
\bibitem[{Dandachi et~al.(2019)Dandachi, De~Domenico, Hoang, and
  Niyato}]{dandachi2019artificial}
\bibinfo{author}{G.~Dandachi}, \bibinfo{author}{A.~De~Domenico},
  \bibinfo{author}{D.~T. Hoang}, \bibinfo{author}{D.~Niyato},
\newblock \bibinfo{title}{An artificial intelligence framework for slice
  deployment and orchestration in 5{G} networks},
\newblock \bibinfo{journal}{IEEE Transactions on Cognitive Communications and
  Networking}  (\bibinfo{year}{2019}).
\bibitem[{Yang et~al.(2019)Yang, Liu, Chen, and Tong}]{yang2019federated}
\bibinfo{author}{Q.~Yang}, \bibinfo{author}{Y.~Liu}, \bibinfo{author}{T.~Chen},
  \bibinfo{author}{Y.~Tong},
\newblock \bibinfo{title}{Federated machine learning: Concept and
  applications},
\newblock \bibinfo{journal}{ACM Transactions on Intelligent Systems and
  Technology (TIST)} \bibinfo{volume}{10} (\bibinfo{year}{2019})
  \bibinfo{pages}{1--19}.
\bibitem[{Li et~al.(2020)Li, Sahu, Talwalkar, and Smith}]{li2020federated}
\bibinfo{author}{T.~Li}, \bibinfo{author}{A.~K. Sahu},
  \bibinfo{author}{A.~Talwalkar}, \bibinfo{author}{V.~Smith},
\newblock \bibinfo{title}{Federated learning: Challenges, methods, and future
  directions},
\newblock \bibinfo{journal}{IEEE Signal Processing Magazine}
  \bibinfo{volume}{37} (\bibinfo{year}{2020}) \bibinfo{pages}{50--60}.
\bibitem[{Smith et~al.(2017)Smith, Chiang, Sanjabi, and
  Talwalkar}]{smith2017federated}
\bibinfo{author}{V.~Smith}, \bibinfo{author}{C.-K. Chiang},
  \bibinfo{author}{M.~Sanjabi}, \bibinfo{author}{A.~S. Talwalkar},
\newblock \bibinfo{title}{Federated multi-task learning},
\newblock in: \bibinfo{booktitle}{Advances in Neural Information Processing
  Systems}, \bibinfo{year}{2017}, pp. \bibinfo{pages}{4424--4434}.
\bibitem[{Paschos et~al.(2020)Paschos, Iosifidis, and Caire}]{paschos2019cache}
\bibinfo{author}{G.~Paschos}, \bibinfo{author}{G.~Iosifidis},
  \bibinfo{author}{G.~Caire},
\newblock \bibinfo{title}{Cache optimization models and algorithms}
  \bibinfo{volume}{16} (\bibinfo{year}{2020}) \bibinfo{pages}{156--345}.
\bibitem[{Li and Avestimehr(2020)}]{li2020coded}
\bibinfo{author}{S.~Li}, \bibinfo{author}{S.~Avestimehr},
\newblock \bibinfo{title}{Coded computing: Mitigating fundamental bottlenecks
  in large-scale distributed computing and machine learning}
  \bibinfo{volume}{17} (\bibinfo{year}{2020}) \bibinfo{pages}{1--148}.
\bibitem[{Fawzi et~al.(2015)Fawzi, Fawzi, and Frossard}]{fawzi2015fundamental}
\bibinfo{author}{A.~Fawzi}, \bibinfo{author}{O.~Fawzi},
  \bibinfo{author}{P.~Frossard},
\newblock \bibinfo{title}{Fundamental limits on adversarial robustness},
\newblock in: \bibinfo{booktitle}{Proc. ICML, Workshop on Deep Learning},
  \bibinfo{year}{2015}.

\end{thebibliography}

\end{document}